%% file: main.tex
\documentclass[twocolumn,trackchanges]{aastex701}%,linenumbers
\usepackage{longtable}
\usepackage{booktabs}
\usepackage[referable]{threeparttablex}

\newcommand{\modify}[1] {{\textcolor{black}{{#1}}}}
\usepackage{pifont}
\usepackage{enumitem}

\newcounter{tabrefcnt}

\begin{document}

\title{A systematic search for long GRBs without supernovae: global properties and rate of events}

\author[0000-0003-0691-6688]{Yu-Han Yang}
\affiliation{Department of Physics, University of Rome ``Tor Vergata'', via della Ricerca Scientifica 1, I-00133 Rome, Italy}
\email[show]{yuhan.yang@roma2.infn.it}

\author[0000-0002-1869-7817]{Eleonora Troja}
\affiliation{Department of Physics, University of Rome ``Tor Vergata'', via della Ricerca Scientifica 1, I-00133 Rome, Italy}
\email[]{eleonora.troja@uniroma2.it}

\author[0000-0002-4129-8195]{Nicolas Monsalves}
\affiliation{Department of Physics, University of Rome ``Tor Vergata'', via della Ricerca Scientifica 1, I-00133 Rome, Italy}
\email[]{Nicolas.Monsalves@uniroma2.it}

\author[0009-0008-9010-2890]{Massine El Kabir}
\affiliation{Department of Physics, University of Rome ``Tor Vergata'', via della Ricerca Scientifica 1, I-00133 Rome, Italy}
\affiliation{Department of Physics, University of Rome ``Sapienza'', P.le Aldo Moro 2, I-00185 Rome, Italy}
\email[]{}

\author[0009-0009-7526-4522]{Narjes Shahamat Dehsorkh}
\affiliation{Department of Physics, University of Rome ``Tor Vergata'', via della Ricerca Scientifica 1, I-00133 Rome, Italy}
\email{}

\author[0009-0001-2089-9899]{Hira Waseem}
\affiliation{Department of Physics, University of Rome ``Tor Vergata'', via della Ricerca Scientifica 1, I-00133 Rome, Italy}
\affiliation{Department of Physics, University of Rome ``Sapienza'', P.le Aldo Moro 2, I-00185 Rome, Italy}
\email{}

\begin{abstract}

Some long duration gamma-ray bursts (LGRBs) are not associated with bright supernovae (SNe), indicating that their explosion mechanism differs from the traditional collapsar channel. 
SN-less LGRBs are rare, thus it remains unclear whether they are sporadic outliers or a substantial component of the GRB population. 
In this work, we use two decades of \textit{Swift} observations to conduct a systematic search for nearby ($z\lesssim$0.35) LGRBs without a confirmed SN association. 
Our analysis reveals a heterogeneous population linked, in part, to star-forming or dusty environments and, in part, to quiescent galaxies. Through the analysis of their high-energy properties, afterglows, and environments, we identify a sample of LGRBs with robust SN limits and distance scales, more than doubling the number of candidate SN-less events. From our fiducial sample, we derive a volumetric rate, not corrected for beaming, of $R = (0.5 \pm 0.2) {\rm~Gpc^{-3}\ yr^{-1}}$, comparable to the apparent rate of short duration GRBs (SGRBs) and standard LGRBs with SNe with luminosities $L \gtrsim\,10^{50}\,{\rm erg\,s^{-1}}$. For plausible beaming factors, the inferred rate of SN-less LGRBs is consistent with several compact binary merger channels, although the allowed parameter space is already narrow if the entire population is attributed to a single progenitor channel. Owing to the small number of candidate events and the uncertainties in their distance scale, our sample leaves the low-luminosity tail unconstrained. A larger sample will be needed to determine whether they are the faint tail of known progenitor channels or evidence of a distinct route to relativistic explosions.

\end{abstract}

\section{Introduction}

Gamma-ray bursts (GRBs) are sudden flashes of high-energy radiation, 
traditionally separated into two phenomenological classes using the observed duration ($T_{90}$) and spectral properties of their prompt gamma-ray emission \citep{Cline1973,Mazets1981,Norris1984,Kouveliotou1993}. 
\citet{Kouveliotou1993} identified a boundary at $T_{90}\simeq2$\,s to distinguish
between long duration GRBs (LGRBs) and short duration GRBs (SGRBs) observed by the Burst and Transient Source Experiment (BATSE). 
This threshold, although instrument-dependent, is still widely used across
multiple missions as an empirical classification criterion.  

Long bursts, i.e. those lasting more than 2\,s, are linked to the deaths of very massive and rapidly rotating stars (collapsars; \citealt{MacFadyen1999}), as demonstrated by their association with broad-lined Type Ic supernovae \citep[SNe; e.g.,][]{Galama1998,Bloom1999,Stanek2003,Hjorth2003,Woosley2006} and their 
preferential location in star-forming regions within their host galaxies, tracing young stellar populations
\citep[e.g.,][]{Bloom2002, Fruchter2006, Savaglio2009}. Short bursts ($T_{90}<2$\,s) are instead associated with neutron star (NS) mergers \citep{Eichler1989}, as established by the joint gravitational wave (GW) and
electromagnetic observations of GW170817/GRB\,170817A \citep{Abbott2017a,Abbott2017b}.  
This evidence has long motivated the use of prompt emission properties as 
a direct probe of the underlying progenitor system, usually interpreted within a two-channel framework of exploding massive stars and NS mergers. 

The large sample of GRBs discovered and localized by the \textit{Neil Gehrels Swift} Observatory (hereafter \textit{Swift}; \citealt{Gehrels2004}) has challenged this simple dichotomy. Several events have shown that the mapping between prompt emission duration and progenitor channel is not one-to-one \citep{Bloom2008,Zhang2009ApJ}.

GRBs 060505 at $z\simeq0.089$ and 060614 at $z\simeq0.125$ were the first cases of nearby LGRBs for which sensitive follow-up observations excluded a bright accompanying SN
\citep{DellaValle2006,Fynbo2006,GalYam2006,Ofek2007}.
Whereas the relatively short duration of GRB\,060505 ($T_{90}\approx$\,4\,s) could 
be reconciled with the tail of the SGRB distribution, GRB\,060614 lasted for $\approx$\,100\,s and was much more
difficult to accommodate within the standard classification scheme \citep{Gehrels2006}.  
Several interpretations were proposed to explain these two SN-less LGRBs, such as a failed collapsar in which little or no radioactive material was ejected \citep{Woosley1993,Fryer1999,Fryer2006,Tominaga2007}, an alternative compact binary progenitor channel involving a white dwarf (WD) and a NS \citep{King2007}
or a black hole \citep[BH;][]{Fryer1999WD,Dong2018}, and the tidal disruption of a star by an intermediate-mass BH \citep{Lu2008,Kiroglu2026}. 
Alternatively, NS mergers with prolonged central engine activity were also considered \citep{Dai2006, GaoFan2006, Rosswog2007, Zhang2007, Metzger2008, Cannizzo2011, Bucciantini2012}. In this interpretation, GRB\,060614 may represent an
extreme case of a SGRB with temporally extended emission \citep[SGRBEE;][]{NorrisBonnell2006}. This hypothesis is supported by its negligible spectral lag, hard initial pulse followed by softer extended emission, and possible evidence of a kilonova \citep{Gehrels2006,Yang2015,Jin2015}. 
The origin of GRB\,060505 remains debated. The burst occurred within a low-metallicity star-forming region of its host galaxy \citep{Thone2014}, consistent with a young ($\lesssim$10 Myr) progenitor, \modify{whereas its association with a kilonova was proposed but not firmly established \citep{Jin2021}}.
%\modify{A thermal component in its early optical afterglow was interpreted as a possible blue KN from an ultra-quick NS merger \citep{Jin2021}, which would reconcile a merger origin with its young environment.}

Another nearby SN-less event, GRB\,111005A \citep{Michalowski2018}, 
associated with the galaxy ESO 580-49 at $z\sim0.013$, exhibited neither a gamma-ray light curve similar to GRB\,060614 nor any evidence of a kilonova. Thus, it could not be entirely excluded that this and other SN-less long bursts were simply the result of a chance projection onto foreground galaxies \citep{Schaefer2006,CobbBailyn2008,Campisi2008}.

More recently, GRB\,211211A provided a decisive example:
although its prompt emission lasted for about a minute, its luminous optical and near-infrared counterpart pointed to an associated kilonova 
at a distance of $\approx$300 Mpc \citep{Troja2022,Rastinejad2022,Yang2022}.
GRB 230307A strengthened this conclusion by showing
that another very bright long duration burst \citep{Sun2025} was accompanied by 
late-time infrared emission consistent with a kilonova \citep{Yang2024,Levan2024}. 
Both bursts occurred outside their galaxy's star-forming regions and showed no evidence of an accompanying SN.

These discoveries cemented the existence of a population of hybrid LGRBs not associated with traditional collapsars and more likely linked to compact binary mergers, as suggested by their kilonova emission. 
However, the observed sample remains small and highly heterogeneous.
The known cases were identified mainly due to their proximity to a nearby
galaxy and subsequent SN limits.  As a result, it is still unclear whether SN-less LGRBs are rare outliers or a substantial component of the  LGRB population, and whether they are all produced by the same progenitor channel.

Systematic searches for this new class of LGRBs were carried out on the larger sample of bursts detected by the \textit{Fermi} Gamma-ray Burst Monitor \citep[GBM;][]{Meegan2009}. \citet{Jiang2023} searched for outliers of the Amati relation \citep{Amati2006} and then examined their spectral lags, hardness ratios, and other prompt properties. 
Their analysis yielded a small sample of candidates (10 out of over 2000 LGRBs). 
A few more candidates were found in searches based on the prompt emission morphology of GRB\,211211A and GRB\,230307A \citep{Wang2025,Tan2025,Zhu2025,Maccary2026}. 
These studies suggest that events like GRB\,211211A are either intrinsically rare or difficult to distinguish from standard LGRBs using prompt emission properties alone.

A search using accurately localized \textit{Swift} bursts is therefore needed to complement the phenomenology of the prompt emission with afterglow properties, galactic environment, and SN constraints.
These diagnostics are critical for assessing the nature of their progenitors
and estimating their event rate. 
\citet{Troja2022} made the first attempt to identify SN-less LGRBs analogous to GRB\,211211A, selecting a sample of 12 nearby ($z\lesssim$0.3) \textit{Swift} LGRBs with no known SN association.
From these, they derived a rate 
of $0.04 \lesssim R \lesssim 0.8~{\rm Gpc^{-3} yr^{-1}}$ (68\% confidence interval), 
implying that LGRBs could represent a non-negligible fraction ($\approx$10\%) of local compact binary mergers. 
A similar event rate was derived by \citet{dong2025statistical}, who performed a statistical comparison of LGRBs with and without an associated SN at different redshifts.

In this work, we build on this approach by carrying out a systematic search for
nearby SN-less LGRBs in the \textit{Swift} sample. We select bursts
within $z<0.35$, and compare their prompt emission, host galaxy, and SN constraints with those of established SGRBs and LGRBs. 
This allows us to assess whether SN-less LGRBs
form a homogeneous population, identify the most promising non-collapsar
candidates, and estimate their contribution to the local rate of compact binary
mergers.
This paper is structured as follows. We detail our sample selection criteria in Sect.~\ref{sec:sample}. Sect.~\ref{sec:analysis} presents our comprehensive analysis of the prompt emission properties, distance scales, host environments, SN/KN limits, and \textit{Swift} detectability. The physical implications and event rates are discussed in Sect.~\ref{sec:discussion}, followed by a summary of our findings in Sect.~\ref{sec:summary}.
Throughout the manuscript we adopt a standard $\Lambda$CDM cosmology \citep{Planck2020} with $H_0$\,$=$\,$67.4$\,km\,s$^{-1}$\,Mpc$^{-1}$, $\Omega_\textrm{m}$\,$=$\,$0.315$, and $\Omega_\Lambda$\,$=$\,$0.685$. 
All magnitudes reported in this work are given in the AB magnitude system.

\section{Sample selection}
\label{sec:sample}

We follow the same approach of \citet{Troja2022} and construct a volume-limited sample of \textit{Swift} GRBs detected and localized by the Burst Alert Telescope \citep[BAT;][]{Barthelmy2005a} between February 2005 and January 2025 and with $z\lesssim0.35$. 
Within this redshift range, meaningful constraints on SN and KN emission can be derived from ground-based follow-up observations. 
At $z\simeq0.35$, the prototypical SN\,1998bw \citep{Galama1998} would peak at an observed magnitude of $i\simeq22$ AB, a depth routinely reached by small- to medium-aperture telescopes. 
Therefore, the absence of a SN at these redshifts can provide a strong constraint on the GRB progenitor.
Searches for KNe are more challenging but still feasible: an AT2017gfo-like kilonova would reach an optical peak of $r\simeq25.5$ mag at $z\simeq0.35$, which can be achieved with large-aperture telescopes.

\begin{figure}[!t]
    \centering
    \includegraphics[width=\linewidth]{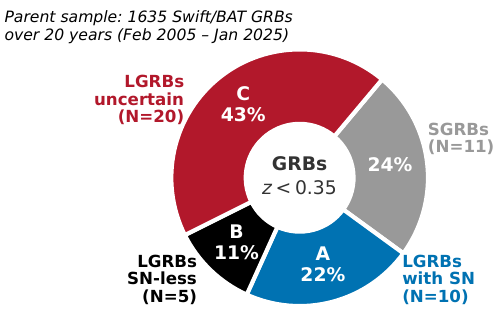}
    \caption{Sample composition for the GRBs at $z<0.35$ considered in this work.
    The sample includes 11 SGRBs (defined as $T_{90}<$2 s), 10 LGRBs with associated SNe (Group A), 5 SN-less LGRBs (Group B), and 20 LGRBs with unconstrained progenitor systems (Group C). Percentages indicate the fractional contribution of each class to the full sample.}
    \label{fig:sample}
\end{figure}

Prompt emission durations were taken from the \textit{Swift}/BAT catalog\footnote{\url{https://swift.gsfc.nasa.gov/results/batgrbcat/index_tables.html}} \citep{Lien2016}. 
We used the observer's frame $T_{90}=2$\,s division as an empirical sorting criterion to distinguish between SGRBs and LGRBs. 
The resulting sample contains 46 GRBs (Fig.~\ref{fig:sample}), whose relevant properties are listed in Tab.~1\ref{tab:sample}:
11 GRBs have $T_{90}<2$ s (SGRBs), the remaining 35 are LGRBs. 
These are divided into three subgroups according to their SN
information: 
10 events have a reported SN association and a confirmed collapsar origin (Group A), 
5 events (GRB 050724, 060505, 060614, 191019A, 211211A) were previously established to lack an accompanying SN down to deep limits (Group B). These bursts have already been extensively studied in the literature and, 
based on their environment or kilonova association, some are
considered likely merger-driven LGRBs.
Our sample does not include GRB\,230307A because it occurred outside of BAT's field of view (FoV).

The last sub-group (Group C) contains the remaining 20 LGRBs with no known SN association.  
These are: GRBs 050219, 050826, 050911, 051109B, 060428B, 
060912A, 061021, 070412, 070521, 080517, 
090417B, 111005A\footnote{This GRB is observationally SN-less \citep{Michalowski2018}. However, given the absence of an X-ray or optical afterglow and the unusual properties of the candidate radio counterpart, we regard its classification as less secure than that of other SN-less GRBs in Group B. Therefore, we conservatively retain the event in Group C.}, 111225A, 130925A, 150424A, 
150727A, 200716C, 200729A, 211227A, and 220611A. 
These bursts ($\approx$40\% of the total sample) define the main sample studied in this work. 
Our aim is to assess whether this subset of nearby LGRBs may arise from an explosion mechanism or progenitor system different than the standard collapsar channel.

\begin{figure*}
    \centering
    \includegraphics[width=0.95\linewidth]{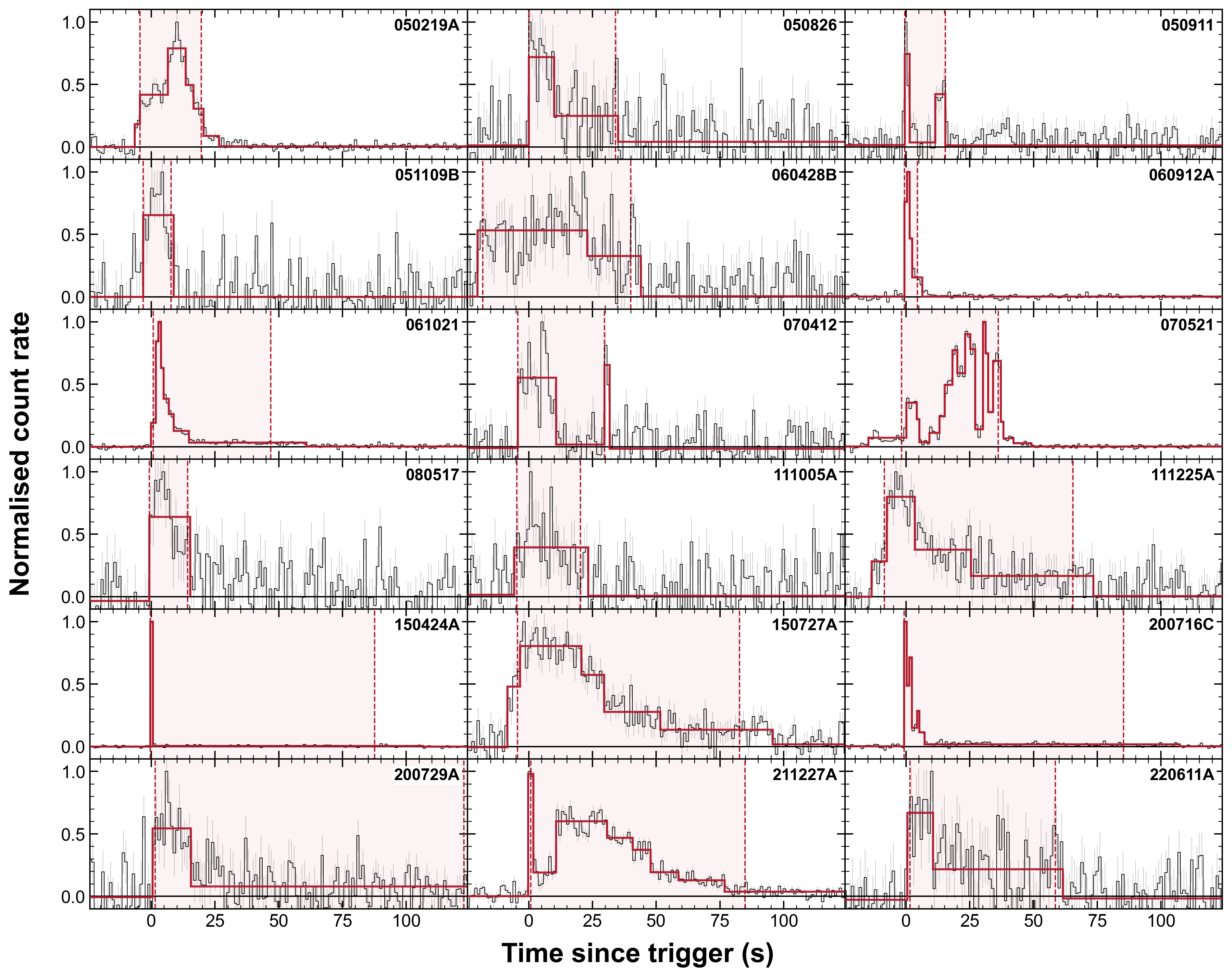}
    \caption{\textit{Swift}/BAT mask-weighted light curves (15--350~keV) of the Group~C sample: 18 LGRBs lacking SN associations as discussed in Sect.~\ref{sec:sample} (ULGRBs 090417B and 130925A are excluded).
    Black step lines show the 1\,s uniformly binned peak-normalised count rate, with the light-gray bars indicating $1\sigma$ uncertainties.
    Overplotted in red are the Bayesian blocks.
    Red dashed vertical lines and the shaded band mark the $T_{90}$ interval. 
    }
    \label{fig:groupC_lightcurves}
\end{figure*}

\input{Tab/sample}

\section{Analysis}\label{sec:analysis}

\begin{figure*}
    \centering
    \includegraphics[width=0.9\linewidth]{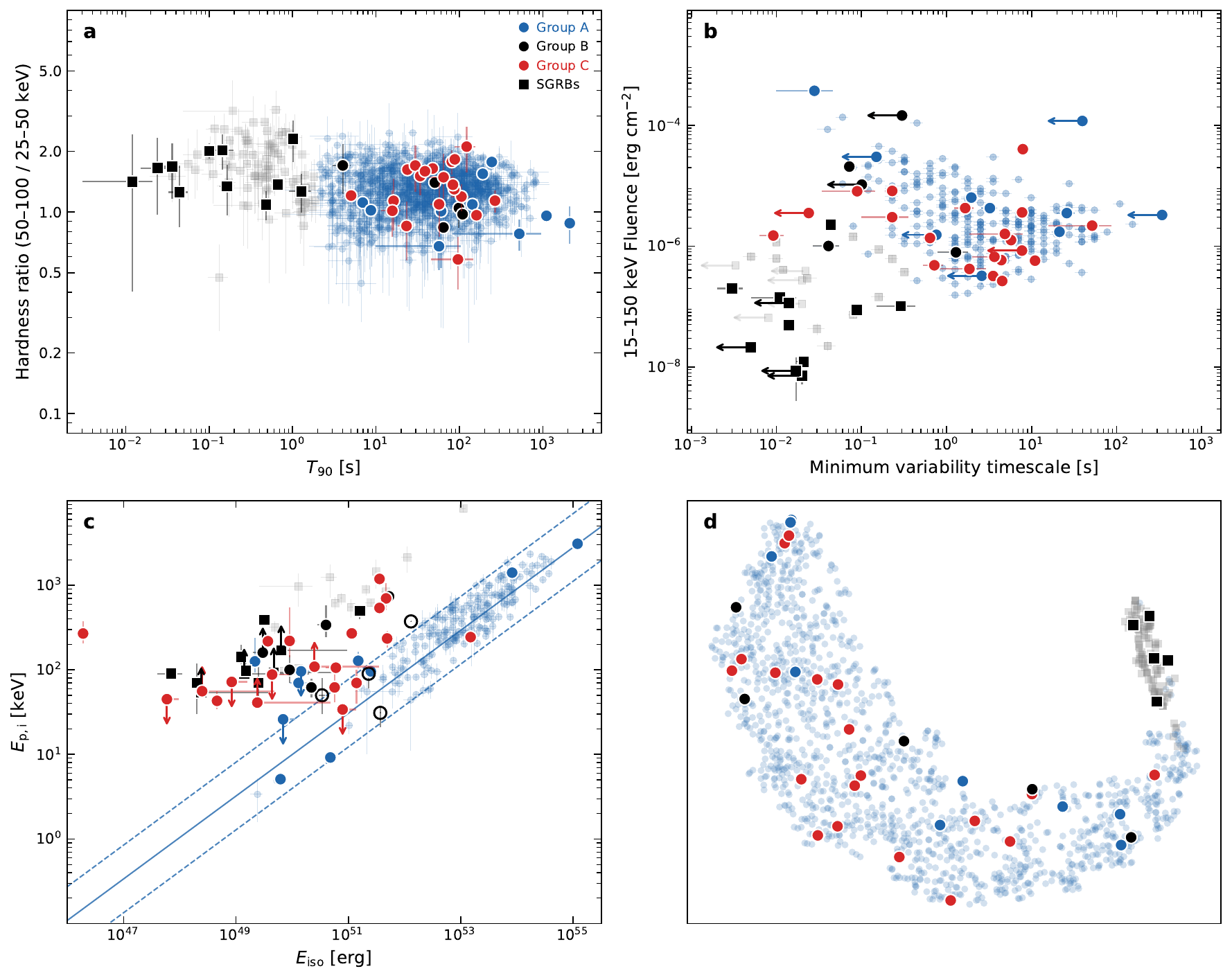}
    \caption{The GRB classification through prompt emission properties: (a) the duration-hardness plane, (b) the variability timescale, (c) the Amati relation and (d) a t-SNE embedding of preprocessed Swift/BAT light curves at perplexity 30.
    For Group B in panel c, the whole-burst measurements (open black circles) lie within or near the LGRB population, whereas the harder main-emission episodes (filled black circles) depart from the canonical Amati locus.
    %The faint GRBs are not included in the t-SNE analysis.
    \textit{An interactive version of this figure is available online, in which hovering over a data point displays the GRB properties; dragging over a region of a panel zooms into it; clicking on a legend entry hides or shows the corresponding group; and selecting an individual burst from the drop-down menus highlights it in all four panels and, for bursts in Group C, displays its redshift track in panel c.}
    }
    \label{fig:classification}
\end{figure*}

\subsection{Prompt emission properties}

In our sample (Group C), most GRBs are dominated by a single prompt pulse 
displaying a variety of durations and temporal profiles (see  Fig.~\ref{fig:groupC_lightcurves}). 
These properties alone do not provide a clear classification. 
The light curve of GRB\,070521 possibly resembles the precursor/short pulse/long tail morphology observed in GRB\,211211A and GRB\,230307A \citep{Dichiara2023}. The light curves of GRB\,061021 and GRB\,200716C are possible examples of a relatively short initial pulse followed by longer lasting low-level emission, whereas GRB\,060912A is an ambiguous case due to its borderline duration $T_{90} \approx$5 s. 
Notable exceptions are the ultralong GRB\,090417B \citep{Holland2010}, GRB\,130925A \citep{Piro2014}, as well as GRB\,150424A and GRB\,211227A, which consist of an initial hard spike followed by a long-lasting tail, resembling the morphology of SGRBEEs \citep{Knust2017, Lu2022}. 
The remaining GRBs in Group C are characterized by a single, continuous emission episode, with Bayesian-block analysis revealing no significant dips or quiescent intervals that would resolve the profile into multiple episodes.

This qualitative inspection of the light curves is complemented by standard quantitative diagnostics of the prompt gamma-ray emission. 
Although the duration/hardness ratio diagram \citep{Kouveliotou1993} remains the dominant classifier, several other properties of the prompt gamma-ray emission are used to refine the GRB classification \citep[cf. Extended Data Fig. 2 in][]{Troja2022}. 
Among them, we consider the minimum variability timescale \citep{Golkhou2014} and the Amati diagram (\citealt{Amati2006}; for alternative schemes, see e.g. \citealt{Lu2010,Minaev2020,Dainotti:2020azn,Nuessle2024,Kang2025}).

Prompt emission properties are primarily taken from the \textit{Swift}/BAT Gamma-Ray Burst Catalog\footnote{\url{https://swift.gsfc.nasa.gov/results/batgrbcat/index_tables.html}} \citep{Lien2016}. 
The durations $T_{90}$ and minimum variability timescales \citep{Golkhou2014} are derived from the mask-weighted 15--350 keV light curves.
When the BAT energy coverage is insufficient to constrain the prompt emission spectral parameters, we incorporate complementary information from other instruments, such as \textit{Fermi}/GBM\footnote{\url{https://heasarc.gsfc.nasa.gov/w3browse/fermi/fermigbrst.html}} and \textit{Konus-Wind}. 
In cases for which only BAT data are available, 
we adopt a Band function model with the photon indices at representative values, $\alpha=-1$ and $\beta=-2.3$ \citep[e.g.,][]{Band1993,Preece2000,Kaneko2006}, and derive the corresponding limits on $E_{\rm p}$ and $E_{\rm iso}$.
Our results are summarized in Tab.~\ref{tab:sample} and Fig.~\ref{fig:classification}.

In the last few years, several attempts have been made to apply machine-learning methods and overcome the limitations of a duration-based classification. 
To date, none of them has provided a robust framework to differentiate between standard LGRBs, SGRBEEs, and SN-less LGRBs. 
Fig.~\ref{fig:classification}d illustrates where our sample falls within the \modify{t-distributed stochastic neighbor embedding \citep[t-SNE;][]{maaten2008}} diagram, generated using the Python package \texttt{ClassiPyGRB}\footnote{\url{https://github.com/KenethGarcia/ClassiPyGRB}}. 
\modify{It shows a two-dimensional representation of the higher dimensional data set built from the \textit{Swift}/BAT light curves in the 15--25, 25--50, 50--100, 100--350 and 15--350 keV bands at 64 ms resolution. 
The relative proximity of GRBs in the map provides information about the similarity of their high-energy properties. 
As in previous works \citep[e.g.,][]{Jespersen2020, Garcia-Cifuentes2023, Dimple2023} the map separates SGRBs from LGRBs: SGRBs concentrate in the right region of the diagram, whereas LGRBs form a structure that starts from the right and extends across the rest of the figure. Groups A, B and C fall on the LGRB structure and spread along it without concentrating in any particular region.}

\subsection{Distance scale}
\label{sec:DisScale}

A critical assumption of our sample is that the selected
bursts lie at $z\lesssim$0.35. 
The redshifts of LGRBs in Group A, for which a SN is detected and spectroscopically confirmed, are considered secure. 
For a few more events, such as GRB 061021 \citep{Fynbo2009} and GRB\,111225A \citep{Thone2014GCN16079}, the distance scale is established by afterglow spectroscopy and is therefore independent of host identification.
Other redshifts rely on the positional coincidence between the GRB and its putative host galaxy. 
As a consequence, some of these associations may be chance alignments with foreground galaxies, leading to an incorrect low redshift assignment. This uncertainty affects most SGRBs and LGRBs in Groups B and C.

We estimate the likelihood of such random alignments in our sample using the formalism of \citet{Bloom2002}, extended by \citet{OConnor2022} to account for XRT localizations.
For each burst, we define $P_{{\rm cc}}$ as the probability that the proposed host 
association is due to a chance alignment. 
To compute this quantity, we use the cumulative number counts of galaxies in the
$r$ band evaluated at the apparent magnitude of the candidate host. Because we are interested in nearby events, which are often hosted in bright galaxies, we adopt the prescription of \citet[][Eq. 3]{Becerra2023} to estimate the galaxies surface density over a broad range of magnitudes (13\,mag\,$\lesssim R\lesssim$\,26\,mag). 
For bright galaxies within 200 Mpc, we estimate the chance coincidence probability through simulations following \citet{Dichiara2020}. 
The derived values are listed in Tab.~\ref{tab:sample}. 

\begin{figure}
    \centering
    \includegraphics[width=\linewidth]{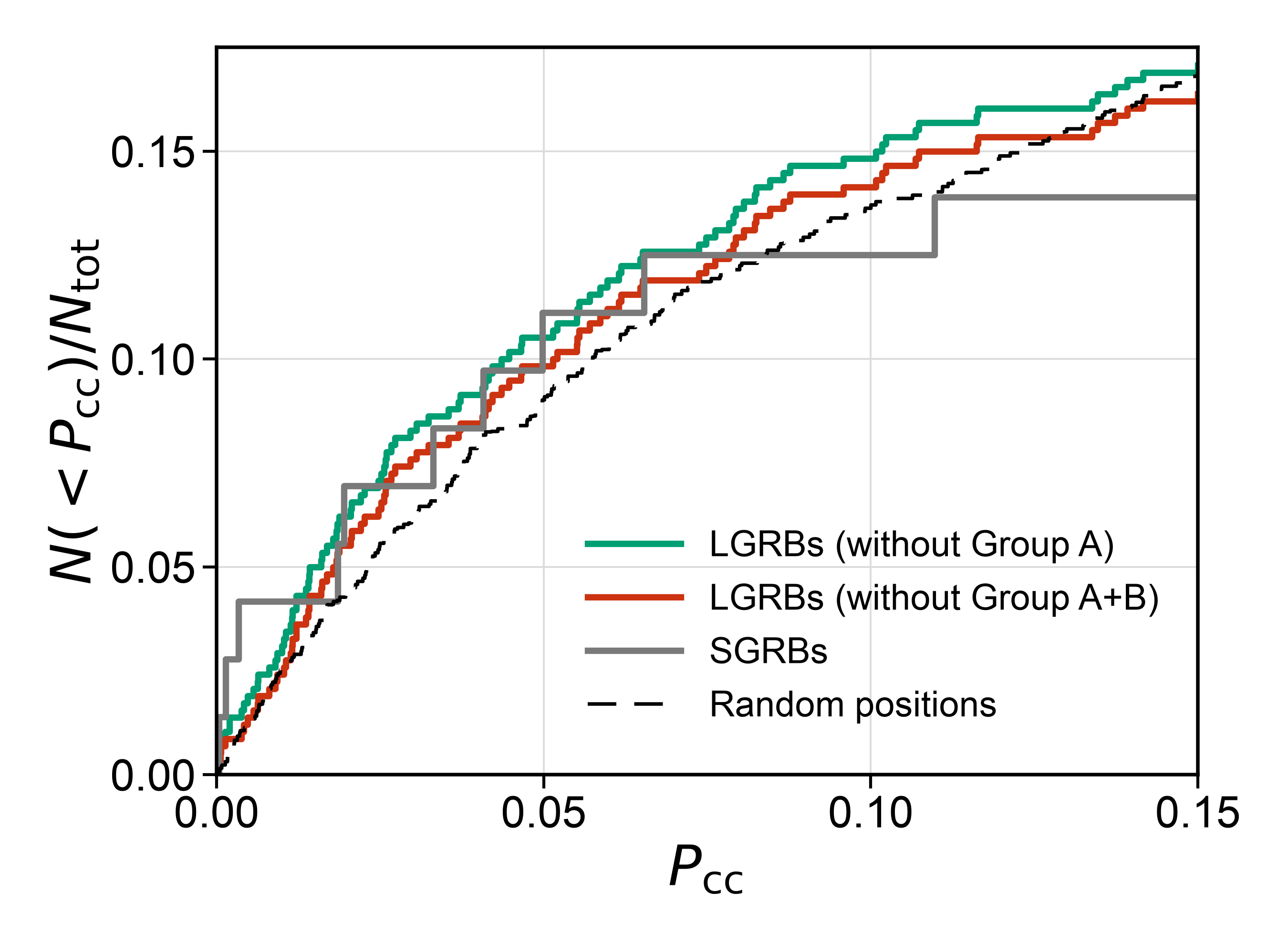}
    \caption{Normalized cumulative distribution of chance coincidence probabilities,
$P_{\rm cc}$, for low redshift ($z\lesssim$0.35) galaxy associations. 
The green curve shows the full \textit{Swift} LGRB sample without direct redshift measurements, whereas the red curve shows the same sample after excluding Group B bursts. The grey curve shows the SGRB distribution. 
The results from random sky positions are shown by the dashed line. 
At low $P_{\rm cc}$, the GRB samples show an excess relative to random positions, indicating that some associations are likely physical. At larger $P_{\rm cc}$, the GRB and random distributions become increasingly similar, suggesting that chance alignments dominate.}
    \label{fig:pcc}
\end{figure}

Overall, the $P_{\rm cc}$ values are small, in most cases below 1\%. 
However, this does not necessarily imply that the probability of chance alignments in the sample is negligible. Since the search is carried out over the full population of $N > 1000$ localized GRBs, even sub-percent probabilities can lead to spurious associations. 
We therefore estimate the contamination by comparing the GRB/galaxy associations to a control sample of random sky positions. 
We generated 10 independent realizations, each comprising 1000 random positions
with errors drawn from the distribution of GRB positional uncertainties.
These were cross-matched with the Legacy Survey catalog \citep{Dey2019}
and associated with a galaxy following the same $P_{cc}$ criteria used for GRBs. We used the available spectroscopic and photometric redshifts to estimate the galaxy's distance scale. 

Fig.~\ref{fig:pcc} shows the cumulative fraction of associations with nearby ($z\lesssim$0.35) galaxies for the sample of \textit{Swift} LGRBs covered by the survey (green solid line), the same sample after removing the well-known SN-less candidates from Group B (red solid line), and the median value of the simulated random positions (dashed line).  
For low  $P_{\rm cc}$ values, 
the LGRBs distribution rises more steeply 
relative to random sky positions, indicative of a higher number of matches. 
The probability that all these associations are the result of random alignments is small ($<$0.3\% assuming Poissonian statistics), confirming that some of these LGRBs are physically associated with low-redshift galaxies. 
After removing the well-known cases discussed in the literature (Group B), an excess remains visible. The probability that this is caused by random associations is $\approx2\%$. 
The size of the observed excess is comparable to our sample size.
We interpret this result as an indication that most ($\approx$15 bursts)  of the proposed low redshift LGRB associations (Group C) are genuine, and that our selection criteria efficiently identified low-redshift GRBs within $z\lesssim$0.35. 
For $P_{\rm cc}\gtrsim3\%$, the LGRB and random position distributions become increasingly similar, implying that associations in this regime are dominated by chance alignments rather than true host identifications.

A similar consideration is valid for the sample of SGRBs (grey solid line in Fig.~\ref{fig:pcc}), for which a probability threshold of 5\% or higher is often considered sufficient for an association \citep{OConnor2022, Fong2022}.  
Based on our simulations of random positions, up to 8\% of the total sample
($\approx$100 SGRBs with an arcsecond position) could yield spurious low-redshift associations with $P_{\rm cc}\lesssim0.05$.  Although the SGRB sample in Tab.~1\ref{tab:sample} was already cleaned of the most likely cases of chance alignments, such as GRB\,061201 \citep{Troja2026} and GRB\,211106A \citep{Laskar2022}, our test suggests that residual spurious associations might still be present in non-negligible numbers.

\

\begin{figure*}
    \centering
    \includegraphics[width=\linewidth]{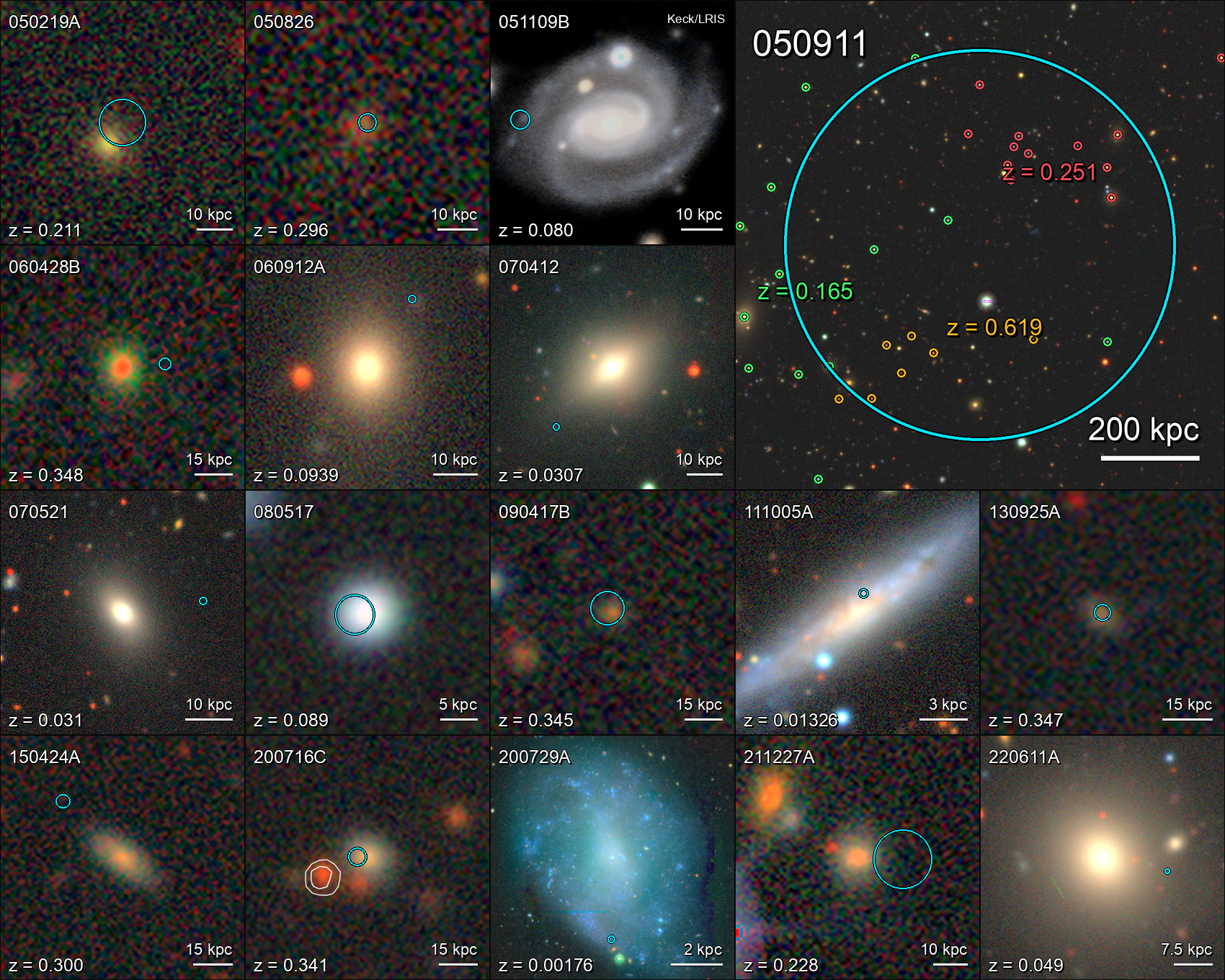}
    \caption{False-color $grz$ cutout images of the LGRBs in Group C from the Legacy Survey \citep{Dey2019}. The image of GRB\,051109B was
    instead acquired with the Keck Low Resolution Imaging Spectrometer (LRIS). 
    The cyan circle marks the GRB localization region. Redshifts are reported in the lower-left corner of each panel, and the physical scale is shown in the lower-right corner. In the field of GRB\,050911, three galaxy clusters are identified, and their members are marked by the colored circles. In the field of GRB\,200716C the white contours show the peak of radio emission. The fields of GRBs\,061021, 111225A, and 150727A are not shown as their faint host galaxies are not visible.}
    \label{fig:hosts}
\end{figure*}

\subsection{Host environment} \label{sec:environment}

\modify{The circumburst and host environment provide an independent diagnostic of the progenitor channel.}
Collapsars are short-lived massive stars still embedded within their progenitor cloud \citep{Reichart2001,Woosley2006}.
Classical LGRBs associated with collapsars tend to reside in low-mass, metal-poor star-forming galaxies \citep{LeFloch2003,Graham2013}, although a fraction of them explode in massive, highly dust-obscured galaxies \citep{Perley2016}.

Compact binary mergers follow a different evolutionary path \citep{Fryer1999}. After the formation of the compact binary, a broad range of delay times and possible natal kicks can move the system away from its birth site before coalescence. 
SGRBs are indeed found in a heterogeneous population of hosts, ranging from early-type galaxies to spirals and actively star-forming systems \citep{OConnor2022,Nugent2022}. 

Moreover, their locations within their parent galaxies are expected to differ: collapsars are preferentially associated with bright UV regions within the host \citep{Bloom2002,Fruchter2006}, while compact object mergers can occur over a broad range of galacto-centric offsets \citep{Fryer1999,Fong2013,OConnor2022}.

\modify{We use the host properties and, where available, the burst offset as additional diagnostics for the likely progenitor channel.}
The candidate host galaxies for the Group C bursts span a wide range of morphologies, including spheroidal systems, interacting galaxies, and large spirals, and therefore do not fit the standard picture of LGRB environments. Their $grz$ images are shown in Fig.~\ref{fig:hosts}.

\begin{figure*}
    \centering
    \includegraphics[width=\linewidth]{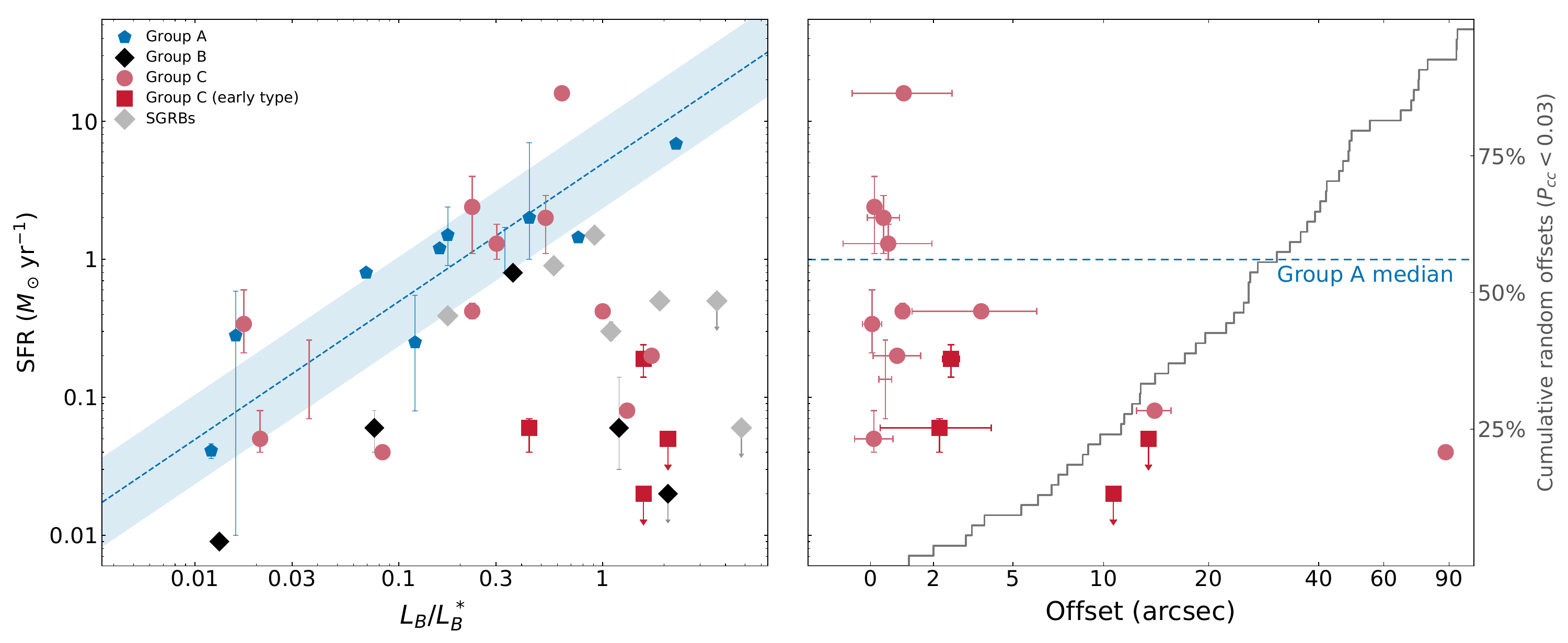}
    \caption{\textit{Left panel:} Star formation rate (SFR) of the candidate host galaxy versus its normalized rest-frame B-band luminosity. The data, compiled from the available literature and this work, are listed in Tab.~1\ref{tab:sfr}. Downward arrows indicate upper limits. Pentagons, black diamonds, circles and squares represent Group A, Group B, Group C and early-type Group C hosts, respectively. The measurements refer to LGRBs in Group C after excluding chance alignments. No SFR measurement is available for the faint host galaxies of GRBs\,111225A and 150727A.
    Light grey diamonds mark SGRBs hosts, only thise with $P_{cc}<3\%$ are reported. The dashed blue line tracks the average
    specific SFR for Group A bursts, and its dispersion is shown by the blue shaded band. SGRBs and most SN-less LGRBs tend to lie below this value.
    \textit{Right panel:} 
    Host galaxy SFR as a function of the angular offset between the burst position and the host center. 
    The horizontal axis uses an inverse hyperbolic sine (asinh) scale. 
    The horizontal dashed line marks the median SFR for galaxies hosting GRB--SN (Group A). The blue solid curve shows the cumulative offset distribution from chance alignments with $P_{\rm cc}<0.03$. 
    \textit{An interactive version of this figure is available online, in which hovering over a data point displays the GRB name, and clicking on a legend entry hides or shows the corresponding burst.}
    }
    \label{fig:sfr}
\end{figure*}

In  Fig.~\ref{fig:sfr}, we compare the host galaxy star-formation rate (SFR) with its rest-frame B-band luminosity $L_B$ (left panel) and with the angular offset between its center and the burst location (right panel).

As found for the broader LGRB population \citep{Christensen2004}, Group A hosts tend to cluster around high values of specific SFR, \(\mathrm{SFR} \approx 6 (L_B/L_B^*)\,M_\odot\,\mathrm{yr}^{-1}\) (dashed line) with an observed \(1\,\sigma\) dispersion of 0.3 dex (shaded area),
where $L_B^*$ is the characteristic luminosity corresponding to $M_B^*=-$20.5 for $z$\,$<$\,0.35 \citep{Tortorelli2020}. 
Group B hosts lie well below this relation (except for GRB\,060505), indicating lower star formation activity per unit B-band luminosity. Group C spans a broader range: some galaxies overlap the Group A relation, other ones occupy the region of SGRBs, whereas the early-type hosts lie in the low SFR corner.

The right panel of Fig.~\ref{fig:sfr} shows that LGRBs in Group C are more strongly concentrated toward smaller angular offsets than the random associations (solid line). The median offset is \(\approx2''\) compared with \(\approx26''\) for the random associations.  
These small offsets support the physical association of most GRBs with their low redshift galaxies. \modify{Conversely, large offset events such as GRBs 051109B, 060912A, and 220611A  occupy a region in which random alignments are common and, therefore, require greater caution.}
The extreme offset of GRB\,200729A is instead less likely to result from a random association.

In many cases, the burst environment supports a compact binary merger progenitor. GRBs\,050219A, 060428B, 060912A, and 220611A are associated with early-type galaxies (red squares in Fig.~\ref{fig:sfr}), while GRB\,050911 is likely associated with a galaxy cluster (although its distance remains uncertain). GRB\,061021 is linked to a quiescent galaxy, and GRB\,211227A occurred in a low-density environment at an offset from its putative host, consistent with a compact binary displaced by a natal kick. By contrast, several bursts are associated with faint blue galaxies (GRBs\,111225A and 150727A) or dusty star-forming environments (GRBs\,080517, 090417B, and 130925A), and display relatively small projected offsets ($\lesssim5$~kpc). Their properties favor young stellar progenitors over neutron star mergers. 

Other peculiar cases occurred in environments that differ from those of typical LGRBs, but do not provide conclusive evidence for a specific progenitor channel. Among them, GRB\,051109B and GRB\,111005A are particularly interesting: within the low-luminosity ($L<10^{49}\,{\rm erg\ s^{-1}}$) subset of our sample, they are the only events for which no underlying faint galaxy has been found. Therefore, they represent the strongest candidates for genuinely low-redshift, low-luminosity GRBs in this group.

Finally, for GRBs\,070412, 070521 and 150424A, the available evidence (e.g. afterglow brightness, or substantial absorption) suggests that the proposed low redshift associations are most likely chance alignments, and we no longer consider them in our analysis.
In all other cases, when no additional information resolves the ambiguity in host association, we retain the galaxy with the lowest $P_{cc}$ as the candidate host.
Below we summarize our detailed findings for each burst in our sample. \smallskip

\textit{GRB\,050219:} The XRT localization of this LGRB intercepts 
a massive ($\approx$10$^{10}$\,M$_{\odot}$) early-type galaxy with little on-going star-formation \citep[$\lesssim$0.06\,M$_{\odot}$\,yr$^{-1}$;][]{Rossi2014}, strongly suggestive of an old stellar population and a merger progenitor. 
A fainter extended source is also detected close to the XRT localization \citep[cf. Figure 1;][]{{Rossi2014}}, although with a higher $P_{\rm cc}$, and could be part of an interacting system \citep[e.g.][]{Kelly2013,Dichiara2026}.  Thus, while the bright early-type galaxy remains the favored association on probabilistic grounds, the GRB birth site is not unambiguous.\smallskip

\textit{GRB\,050826:} The burst is located within a compact galaxy with a rest-frame $B$-band luminosity of $\sim0.3L_\star$, and a moderate star-formation rate of $\sim3\,M_\odot\,{\rm yr}^{-1}$ \citep{Mirabal2007}. 
Although the galaxy's stellar mass is rather high,  $M_\star\sim1.3\times10^{10}\,M_\odot$, 
its relatively low metallicity, $12+\log({\rm O/H})\lesssim8.4$, and dust content, $E(B-V) \approx 0.6$, point to a collapsar progenitor \citep{Levesque2010, 2021A&A...656A.136T}. \smallskip

\textit{GRB\,050911:}  The BAT localization intercepts the galaxy cluster EDCC~493, whose members have redshifts $z\simeq0.165$ \citep{Berger2007}. Within the error region
we identify two additional clusters at $z\sim0.251$ and $z\sim0.619$, respectively \citep{Wen2024}. These systems are less massive than EDCC~493 but have a smaller angular extent, yielding overall similar probabilities of association with GRB\,050911. 
The cluster environment, rapid afterglow fading, and prompt emission make GRB\,050911 compatible with a compact object merger \citep{Page2006}. However, the lack of a secure afterglow position prevents a unique cluster assignment, and the burst distance scale remains uncertain. \smallskip

\textit{GRB\,051109B:}  The XRT localization of this LGRB lies close to a bright barred spiral galaxy at $z=0.080$.  The burst position lies on a faint outer spiral arm, where Keck/LRIS spectroscopy revealed emission lines from an actively star-forming knot \citep{Perley2006GCN}. The local environment makes the GRB site compatible with recent star formation but do not uniquely identify the progenitor channel. \smallskip

\textit{GRB\,060428B:} The burst lies only $\simeq2.6''$ from the center of a bright red elliptical galaxy at $z=0.348$, whose optical spectrum is dominated by an old stellar population, with a strong 4000~\AA\ break and little star formation \citep{Perley2007AIP}.  Such an environment would be compatible with a compact binary merger origin.  However, deep Keck imaging also revealed a compact blue source coincident with the afterglow position, with $P_{\rm cc}\approx1.5\%$, raising the possibility that the red galaxy is a foreground object rather than the true host.  Therefore, the main uncertainty for this event is the association between the GRB and the galaxy itself (Sect.~\ref{sec:DisScale}). \smallskip

\textit{GRB\,060912A:}  Its proximity to a nearby ($z\sim$0.0936) elliptical galaxy with SFR$\lesssim0.02\,M_{\odot}$\,yr$^{-1}$ suggests a possible compact binary origin. 
Deep optical imaging shows a distant ($z\approx0.9$) galaxy coincident with the afterglow position \citep{Levan2007}, initially considered a more likely host for a LGRB despite its higher $P_{\rm cc}$. 
However, the burst intermediate duration ($T_{90}\simeq5$ s) is not a robust indicator of a collapsar origin: \citet{Bromberg2013} estimated that $\approx$10\% of the non-collapsar sample of \textit{Swift} bursts may last longer than 3 s. 
Its red afterglow suggests moderate dust extinction along the sightline ($A_V \approx$\,0.2\,mag) but does not break the degeneracy in the distance scale.
Therefore, in this case as well, \modify{the main uncertainty in the GRB classification is the identification of the true host galaxy} (Sect.~\ref{sec:DisScale}). \smallskip

\textit{GRB\,061021:}
The host galaxy is faint with $R\approx24.5$~AB, corresponding to a rest-frame $B$-band luminosity $L_B \sim 0.02-0.03\,L_*$, and is characterized by modest extinction, $E(B-V)$\,$\approx$0.1, low stellar mass, $M_\star$\,$\sim$\,$3\times10^{8}\,M_\odot$, and low star-formation rate, $\sim$\,$0.05\,M_{\odot}$\,yr$^{-1}$\citep{Kruhler2015,Vergani2015}. The spectral energy distribution (SED) displays a prominent 4000 break \citep{Vergani2015}, indicative of an old stellar population. 
Its low specific star-formation disfavors a collapsar explosion, although its low mass and low luminosity are atypical for SGRB hosts \citep[cf. Fig. 14;][]{OConnor2022}. 
\smallskip

\textit{GRB\,070412:} The burst is located in the outer halo of a bright active galaxy, within an overdensity of galaxies at $z$=0.037. The galaxy's spectrum shows a strong 4000~\AA\ break and absorption lines, indicative of an old stellar population with little on-going star-formation, $\lesssim$\,$0.1\,M_{\odot}$\,yr$^{-1}$.
Optical imaging of the field identifies a faint ($I\approx$24.6 mag) unresolved source within the XRT position. 
Despite its bright X-ray afterglow, deep limits on the optical counterpart were derived ($R\gtrsim 24$ mag at $\approx$1 hr; \citealt{Malesani2007GCN}). This implies either a high redshift origin or
 large dust extinction, which is not consistent
with the burst location in a halo environment. Based on these considerations, we flag this event as a chance alignment. 
\smallskip

\textit{GRB\,070521:} Another example of LGRB in the outskirts of a 
bright and passive galaxy. Several fainter galaxies were identified close to or within the XRT localization \citep{Perley2009}, however, on probabilistic grounds, they are less favored. 
The afterglow properties are a key discriminant: the large $N_H$ value inferred from X-ray spectroscopy and the high X-ray to optical flux ratio point to a dusty local environment, which is not compatible with the low-redshift association. Based on these considerations, we flag this event as a chance alignment.  \smallskip

\textit{GRB\,080517:} The XRT localization of this LGRB intercepts a compact and relatively massive, metal-rich galaxy, characterized by an evolved stellar population as well as recent episodes of star-formation with SFR$\approx8-16\,M_{\odot}$\,yr$^{-1}$ \citep{Stanway2015}. 
Such an environment could harbor both massive star progenitors and older compact binary systems, 
although it more closely resembles the sample of dark LGRBs. \smallskip

\textit{GRB\,090417B:} This burst occurred along a very dusty sightline ($A_V \gtrsim$10; \citealt{Holland2010}). The inferred SFR is $\approx1.3\,M_{\odot}$\,yr$^{-1}$ \citep{Niino2017}. The high dust content, stellar mass, and metallicity are atypical for optically-selected LGRBs, but fit within the sample of dark LGRBs. Such a high extinction is rarely seen in SGRBs \citep[cf. Figure 3 in][]{Troja2023}. \smallskip

\textit{GRB\,111005A:} This burst lacks an X-ray and optical afterglow, and was localized through its candidate radio counterpart. This is associated with ESO 580-49, a moderately star-forming (SFR$\sim0.4\,M_{\odot}$\,yr$^{-1}$) galaxy viewed edge-on. 
Significant extinction, $A_V \approx 2$ mag, was measured along the GRB sightline. However, the position of the radio source is located within a metal-rich environment with little ongoing star formation \citep{Tanga2018}.  
Although the global properties of the host galaxy are not conclusive, the characterization of the local environment tends to favor a non-collapsar origin. 
A caveat to this interpretation is that it relies on the association between the GRB and the candidate radio counterpart.  \smallskip

\textit{GRB\,111225A:} 
The host galaxy is faint with $R\approx24$\,mag \citep{Niino2017}, corresponding to a rest-frame $B$-band luminosity $L_B \sim 0.03\,L_*$ (Tab.~\ref{tab:sfr}). 
Optical spectroscopy \modify{(Fig.~\ref{fig:111225Ahostspec})} reveals the presence of nebular emission, indicating on-going star formation, \modify{from which we derive ${\rm SFR}\gtrsim0.07\,M_\odot$\,yr$^{-1}$ \citep{Kennicutt2012}. 
Modeling of the galaxy's SED yields ${\rm SFR}\approx0.26\,M_\odot$\,yr$^{-1}$ and $E(B-V)\approx0.22$ \citep{Taggart2021}, suggesting significant dust-obscured star-formation.}
The environment appears broadly consistent with a massive star progenitor \citep{Niino2017}.
\smallskip  

\textit{GRB\,130925A:} This burst is located 0.12\arcsec~($\sim600$\,pc) from the nucleus of a young, star-forming ($\approx3\,M_{\odot}$\,yr$^{-1}$), edge-on spiral galaxy \citep{Schady2015}. Its dusty sightline \citep[$A_V\approx5-8$\,mag; ][]{Greiner2014} and metal-rich ($\sim 1.5Z_\odot$) birth site
are not typical of LGRB or SGRB environments. \smallskip

\textit{GRB\,150424A:} The GRB is $\approx$7\arcsec north east of a bright spiral galaxy ($P_{\rm cc}\sim$0.02) with a spectroscopic redshift $z\simeq$0.300, and only 0.4\arcsec from a fainter galaxy ($P_{\rm cc}\sim$0.06) at $z\approx$1.2 \citep{Jin2018}. 
The afterglow properties offer a way to break the degeneracy: \citet{Knust2017} presented marginal evidence for a dip in the UV flux, which could be interpreted as a Lyman break at $z\approx1$. The bright X-ray afterglow also disfavors the association with the low-redshift galaxy, which would imply a large offset and presumably a rarefied circumburst environment. 
For this GRB, we derive an X-ray flux to gamma-ray fluence ratio of log\,$F_X/S_{\gamma}$\,$\approx$-6, consistent with a modest offset. 
Based on these considerations, we flag this event as a likely chance alignment with a low redshift galaxy.  \smallskip

\textit{GRB\,150727A:} The information on this GRB host derives from the VLT/XSHooter spectrum, briefly discussed in \citet{Selsing2019} \modify{and re-analyzed in this work to derive its basic properties. 
From the strenght of the H$\alpha$ line, we infer ${\rm SFR}=0.34^{+0.26}_{-0.13}\,M_\odot$\,yr$^{-1}$ \citep{Chabrier2003,Kennicutt2012} after correcting for intrinsic  extinction with $E(B-V)=0.22^{+0.24}_{-0.21}$\,mag. The latter value is derived from the Balmer decrement H$\alpha$/H$\beta=3.5^{+1.0}_{-0.7}$ assuming a \citet{Cardelli1989} extinction law. 
In late-time GROND imaging, the host galaxy is detected at $r'=24.5\pm0.2$\,mag (corrected for Galactic extinction), unresolved and $\approx0.05$\arcsec\ from the afterglow position. Its brigthness corresponds to a rest-frame  $L_B\approx0.017\,L_B^*$, and implies a relatively high specific SFR. 
A faint star-forming galaxy with modest amount of dust and no measurable offset from the afterglow position is typical of LGRB hosts and favors a massive star progenitor.}\smallskip

\textit{GRB\,200716C:} the optical counterpart is coincident with a galaxy
at $z=0.341\pm0.004$ \citep{Giarratana2023}. 
\modify{This GRB was proposed as a candidate lensed burst at higher redshift \citep{Wang2021,Yang2021}. However, subsequent analysis 
found that it is not a compelling case \citep{Mukherjee2022, Zhao2025}. 
Therefore, we consider the nearby galaxy as its putative host.}
Weak emission
lines in the optical spectrum imply only a modest star-formation rate,
${\rm SFR}=8.8\times10^{-42}L_{\rm [O\,II]}\simeq0.2\,M_\odot\,{\rm yr}^{-1}$.  
The detection of diffuse radio emission was interpreted by \citet{Giarratana2023} as evidence of high star formation, $\approx300\,M_\odot\,{\rm yr}^{-1}$. Reconciling the optical and radio estimates would require a dusty line of sight with $A_V\gtrsim5$ mag, which is not supported by the GRB afterglow color.  \modify{An alternative interpretation is, therefore, that the radio emission is not dominated by star formation at the GRB site, but is associated with the unrelated nearby sources visible in Fig.~\ref{fig:hosts}.} \smallskip

\textit{GRB\,200729A:} The XRT localization overlaps with the nearby ($\approx$5.5 Mpc) galaxy NGC~4242, a face-on unbarred spiral galaxy with little on-going star formation 
$\lesssim\,0.09\,M_\odot\,{\rm yr}^{-1}$ \citep{Lee2011}. 
Assuming a physical association with this local galaxy yields a sub-energetic burst with $E_{\gamma,\rm iso} \sim 10^{46}$\,erg. 
\modify{However, pre-explosion imaging reveals a faint and red galaxy within the X-ray error circle \citep{2020GCN}. If this source is the true host, the burst could be located at $z\gtrsim1$.}
A higher redshift would imply more standard energetics, although the burst would still deviate from the Amati relation for LGRBs. \smallskip

\textit{GRB\,211227A:} The putative host galaxy lies outside the XRT error circle, and its global properties are consistent with those of field galaxies, as observed for many SGRB hosts \citep{Ferro2023}.
Our analysis of archival Gemini observations (see Appendix) reveals no additional candidate host 
within the XRT localization down to $i\gtrsim26.2$ mag. 
We derive an X-ray flux to gamma-ray fluence ratio of log\,$F_X/S_{\gamma}$\,$\approx-$8.1, which 
is similar to GRB\,211211A and GRB\,230307A \citep{Yang2024} and supports its
localization in a low density medium, outside of  the low redshift galaxy.  
The environment and large offset point to a merger origin.

\smallskip

\textit{GRB\,220611A:} The burst lies within the light of a bright ($r\approx$14.7 mag) lenticular galaxy, MCG-06-10-007, at $z\approx0.049$. The optical spectrum shows no signs of ongoing star-formation, 
SFR $\lesssim$0.05\,$M_\odot\,{\rm yr}^{-1}$. 
\modify{A fainter distant ($z\approx2.4$) star-forming galaxy is detected underlying the GRB position.  The former would be consistent with a merger progenitor, the latter with an ordinary LGRB from a collapsar. Similarly to GRBs\,060912A, 060428A, and 200729A, the interpretation of this GRB strongly relies on the identification of its true host galaxy. }

\subsection{Constraints on the supernova/kilonova emission}
\label{sec:sn-kn-constraints}

The most direct way to distinguish between massive star and merger
progenitors is to search for the thermal transients expected to accompany each
channel. The core-collapse of a massive star is expected to produce a bright SN powered by the radioactive decay of freshly synthesized $^{56}$Ni, and indeed many LGRBs (Group A) are associated with broad-lined Type Ic SN 
\citep[e.g.][]{Pian2006,Bufano2012,Modjaz2016}. 

A NS merger can produce a rapidly evolving kilonova powered by the radioactive decay of neutron-rich $r$-process material \citep{LiPaczynski1998}. Detecting either component therefore provides strong evidence for the physical origin of the burst, while deep non-detections can rule out ordinary 
GRB-SNe \citep[e.g.][]{Fynbo2006} or constrain the amount of ejecta from a merger \citep[e.g.][]{OConnor2020, Troja2023}. 
For nearby LGRBs (Group C), where such transients should be
observable with sensitive follow-up, the presence or absence of SNe and KNe is
therefore a critical diagnostic for identifying their progenitors and for
separating them from classical LGRBs and chance associations with
foreground galaxies.

\begin{figure}
\centering
\includegraphics[width=\linewidth]{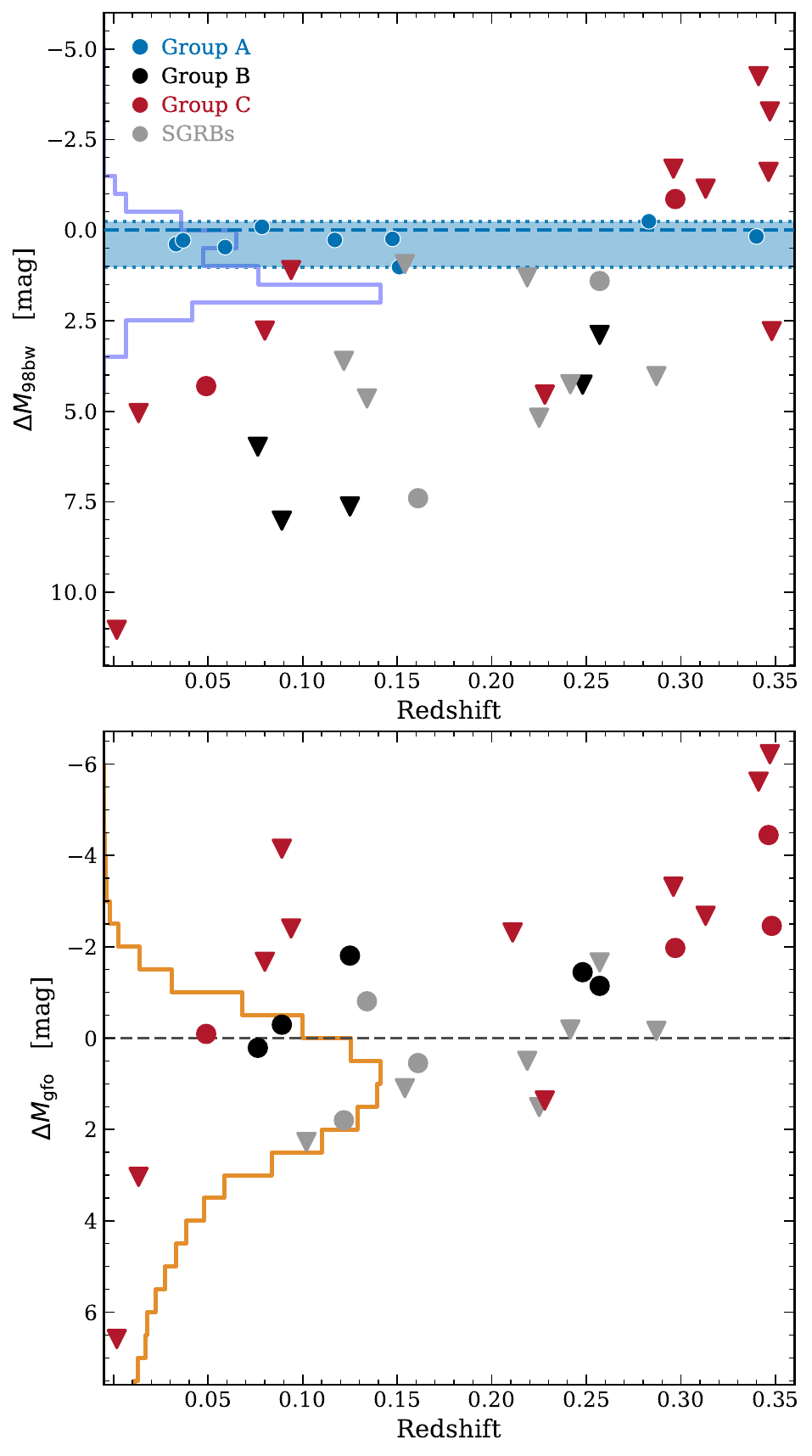}
\caption{ Constraints on the supernova (top) and kilonova (bottom) emission for our sample of nearby GRBs. 
Transient measurements (circles: detections; downward triangles: upper limits) are compared to the template lightcurves of SN1998bw and AT2017gfo. 
$\Delta M$ is the difference between the observed measurement and the template at the same rest-frame epoch and at the same rest-frame wavelength.
For each burst, only the tightest constraint is reported, and the corresponding measurements are listed in Tab.~1\ref{tab:constraints}.
The blue shaded band shows the dispersion of Group~A bursts. 
%\textbf{, for which $\Delta M$ is calculated at the SN peak}. 
The histogram reports the relative brightness distribution for a sample of stripped-envelope SNe from \citet{Prentice2016} in the upper panel and for the
kilonova model grid of \citet{Wollaeger2021} in the lower.}
\label{fig:SNKN}
\end{figure}

We compare the available follow-up observations for the sample with two reference transients, representing the prototypical GRB-associated SN \citep[GRB\,980425/SN1998bw;][]{Galama1998, Clocchiatti2011} and the best-sampled KN \citep[GRB\,170817A/AT2017gfo; e.g., ][]{Coulter2017, Abbott2017b}, respectively.
The comparison was carried out by shifting the two reference light curves to the burst redshift and matching them to the observed filter using the k-correction 
derived from their spectral energy distributions.
Measurements were corrected for Galactic extinction 
\citep{Schlafly2011} using the extinction law of \citet{Fitzpatrick1999}
with $R_V=3.1$.

Interpreting these constraints requires an estimate of the dust extinction along the line of sight, as an accompanying SN could otherwise be obscured. When not available from the literature, we estimated the intrinsic extinction $A_{V,z}$ by comparing the optical
afterglow spectral energy distribution with the X-ray continuum from the XRT online repository \citep{Evans2009}, assuming standard closure relations to constrain their spectral indices 
($\beta_o=\beta_X$ or $\beta_o=\beta_X-0.5$; \citealt{Jakobsson2005}). 
Although the optical emission often falls below the power-law extrapolation of the X-ray spectrum, the available observations are generally too sparse to distinguish dust extinction from spectral curvature associated with a synchrotron cooling break. In those cases, we report an upper limit to the intrinsic extinction derived for $\beta_o = \beta_X$. 
The resulting values of $A_{V,z}$, listed in Tab.~1\ref{tab:sample}, were used to correct the GRB observations.  

For each GRB, we select the tightest limit on the associated SN emission and plot it in Fig. \ref{fig:SNKN} (top panel), whereas the tightest limit on the KN is presented in  Fig. \ref{fig:SNKN} (bottom panel). The corresponding measurements are listed in Tab.~\ref{tab:constraints}, and the method is detailed in Appendix~\ref{app:snkn}.

\subsection{Detectability}\label{sec:detect}

\begin{figure*}
    \centering
    \includegraphics[width=\linewidth]{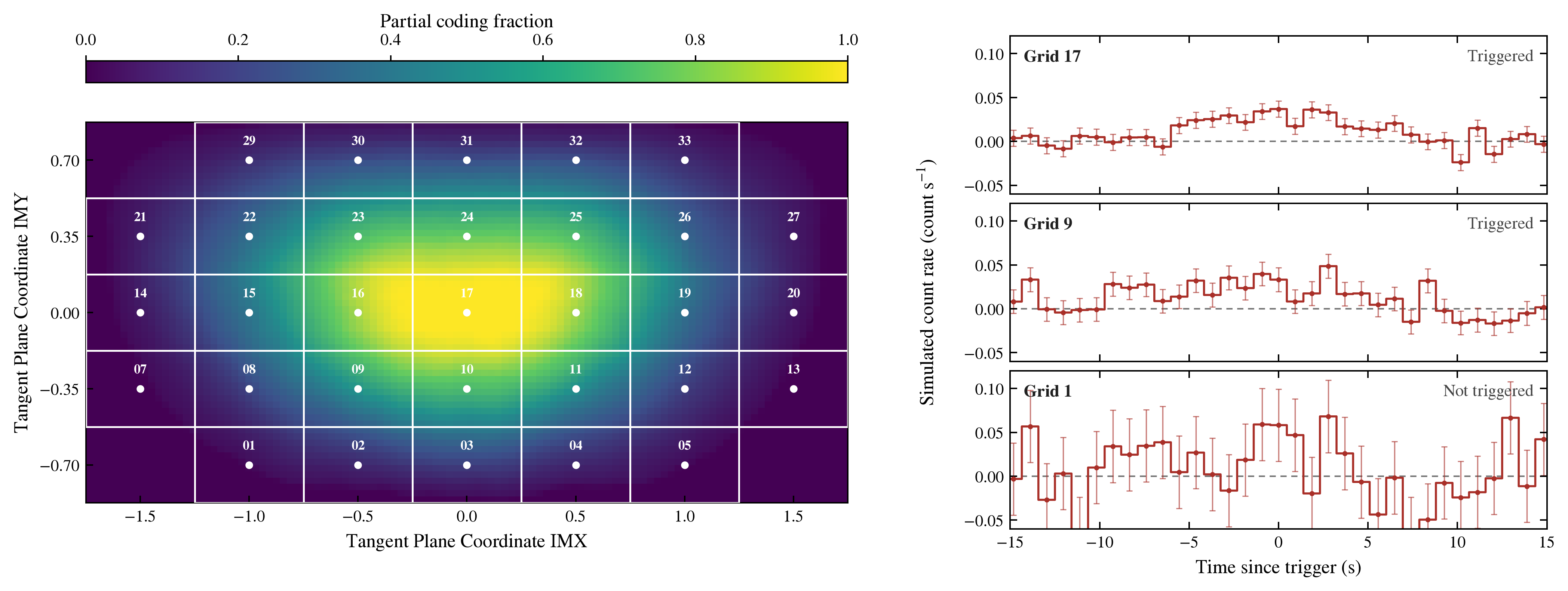}
    \caption{\textit{Left:} BAT partial coding map and response grid, as defined in  \citet{Lien2014}. 
    \textit{Right:} Simulated light curves of GRB~051109B at $z=0.10$ for grid IDs 17, 9, and 1, corresponding to partial coding 100\%, 50\%, and 4\%, respectively. The light curves are shown in the 15--350 keV band, assuming 27,000 enabled detectors. }
    \label{fig:batsim}
\end{figure*}

We defined a volume limited sample of bursts within $z_{\rm max}=0.35$.
However, not every burst would have been detectable throughout this volume. 
Therefore, to convert the observed number of bursts into a volumetric rate, we need to derive the redshift range over which each event could have been detected by BAT and subsequently localized by the XRT.

The BAT trigger threshold depends on the intrinsic luminosity, spectrum, and temporal structure of each burst, as well as its position within the BAT FoV
and the number of active detectors at the time of observation \citep{Lien2014}. Moreover, the identification of the GRB host galaxy requires at least an X-ray localization, introducing additional selection effects due to afterglow brightness and observing constraints. 
In our calculations we model these sources of incompleteness, which can cause low-redshift bursts to be missed.

We assign to each $i$-th burst an effective volume, \(V_i\), obtained by determining its joint BAT and XRT detectability as a function of redshift and mission epoch.
Its value is calculated as: 
\begin{equation}
V_i = \frac{1}{\Omega T}
\int_0^{T} 
\int_0^{z_{\rm max}}
\frac{\Omega_{i}(t,z) \eta_{X,i} (t,z)}{1+z}
\frac{dV_c}{dz}\,dz\,dt
\end{equation}\\
where $\Omega$\,$\approx$\,2.2 sr is the BAT FoV, 
$T\simeq20$ yr is the mission lifetime, 
$dV_c$ is the comoving volume element within $z_{\rm max}=0.35$, and $\eta_{X,i} (t,z)$ is the X-ray afterglow detection efficiency. 

We assumed negligible change in the XRT sensitivity over time, that is $\eta_{X,i} (t,z) \approx \eta_{X,i} (z)$, and an average slew efficiency of $\approx$85\%, that is $\eta_{\rm MAX} \approx 0.85$. 
Its value was estimated by shifting the observed XRT afterglow light curve of each burst to redshift \(z\). It was set to $\eta_{\rm MAX}$ when the predicted afterglow flux exceeded the XRT detection threshold and zero otherwise.

We defined $\Omega_{i}(t,z)$ as the effective BAT solid angle over which burst $i$ would trigger at redshift $z$ and mission epoch $t$.
It depends on the burst luminosity and redshift: faint bursts are detectable only within the most sensitive, fully coded region and therefore sample a smaller solid angle than bright bursts, which can be detected over a larger fraction of the detector plane. 
It also evolves with time as the number of active BAT detectors decreased from $\approx$30,000 in 2005 to $\approx$16,000 in 2025, progressively reducing the instrumental sensitivity.

To approximate the \textit{Swift}/BAT instrumental response, we used the trigger simulator developed by \citet{Lien2014}, as implemented in the \texttt{simmes} package \citep{Moss2022}. 

To estimate how the burst would appear at different redshifts, we first transformed the observed light curve to the source frame and subsequently placed the resulting template at different luminosity distances, logarithmically spaced between 50 Mpc and 2 Gpc. 
Stochastic source and background fluctuations were added to each simulated light curve. 
An important caveat of this approach is that the input template contains only the emission detected in the original BAT observation. Moving this template to a lower redshift brightens the measured signal but cannot recover intrinsically faint components that have fallen below the detection threshold at the original distance. Thus, for \(z<z_{\rm obs}\), the true detection rate may be underestimated if prompt emission components are absent from the input light curve.

For each redshift, we drew 31 positions within the BAT response grid (Fig.~\ref{fig:batsim}) and weighted them by their solid angle, such that the simulations represented sources distributed uniformly on the sky rather than on the detector plane. 
Within the coded region, the solid angle element $d\Omega$ was calculated from the tangent plane coordinates IMX and IMY as: 
\begin{equation}
    \Delta \Omega =    \frac{\Delta \mathrm{IMX}\,\Delta \mathrm{IMY}}   {\left(1+\mathrm{IMX}^{2}+\mathrm{IMY}^{2}\right)^{3/2}}.
\end{equation}

 We also used the HEASoft task \texttt{batmaskwtimg} to calculate the relative contribution of the four BAT detector quadrants required by the trigger algorithms \citep{Lien2014}.

For every combination of redshift and detector position, we generated 10 independent realizations, corresponding to 310 trials at each redshift. Each realization was passed through the BAT trigger criteria implemented in \texttt{simmes}. A burst was classified as detected when at least one trigger criterion was satisfied, and its detection probability was quantified as the ratio between successful triggers and trials. 
An example of simulated light curves is shown in 
Fig.~\ref{fig:batsim} (right panel) for different
grid positions.

\section{Discussion}\label{sec:discussion}

\subsection{Non-collapsar LGRBs}\label{sec:LGRBs}

In Sect.~\ref{sec:analysis}, we examined the properties of a sample of nearby LGRBs with no known SN association. We considered their prompt emission, their environment, their limits on supernova and kilonova emission, and assessed whether their distance scale could be underestimated because of chance alignments with foreground galaxies. 
Our main results can be summarized as follows. 
\begin{enumerate}[label=\Roman*.,topsep=0pt, parsep=0pt, , itemsep=0pt,
leftmargin=15pt]
\item  The prompt emission by itself does not solve the ambiguity in classification, but it helps identify outliers for further study (Fig.~\ref{fig:classification}). 
Three bursts (GRBs\,150424A, 200716C, and 211227A) are characterized by minimum variability timescales $\lesssim$0.1 s, and a light curve morphology consistent with SGRBEEs (see Fig.~\ref{fig:groupC_lightcurves}). 
Four more events (GRB\,060912, 061021, 070412, and 070521) have variability timescales $\lesssim$1 s,  at the lower end of the LGRB distribution.

Half of the Group C bursts are outliers
of the Amati relation (Fig.~\ref{fig:classification}, panel $c$). Only in some of these cases would a wrong redshift assignment help explain the inconsistency. In other cases, the bursts remain outliers regardless of the distance scale. These are GRBs\,061021, 150424A, 200716C, and 200729A.

\item For high quality associations ($P_{\rm cc}$\,$\lesssim$3\%), the sample is not dominated by chance alignments. 
At low $P_{\rm cc}$ the observed sample shows an
excess over random associations (Fig.~\ref{fig:pcc}), and the probability that all such associations are accidental is very small: $\lesssim$0.3\%
for Group B+C bursts,  $\lesssim$2\% for Group C bursts. We estimate $\approx$5 possible spurious associations within Group C. Therefore, the majority of bursts in this sample form an interesting population of low-redshift LGRBs whose progenitors remain ambiguous. 

\item The environment provides the strongest evidence that the population of nearby LGRBs is not homogeneous (Sect.~\ref{sec:environment}). 
Only a minority of events are hosted by
galaxies resembling the ordinary LGRB population. Other bursts are still
likely tied to massive stars but occur in dusty, massive, or high-metallicity
environments. 
\modify{A sizeable subset of bursts (GRBs 050219A, 050911, 060428B, 060912A, 061021,
200716C, 211227A, and 220611A) has host morphologies, low star-formation
rates, and/or galactocentric offsets that make a merger progenitor plausible. This conclusion holds also when the less secure host associations (e.g. GRB\,060912A, GRB\,220611A) are removed.}

Our analysis leads us to exclude four bursts (GRBs\,050911, 070412, 070521, and 150424A) 
from Group C due to their uncertain distance scale.

\item The SN constraints for bursts in Group C are generally weaker than those for SGRBs and Group B. The available observations exclude a luminous SN1998bw-like event for approximately one-third of Group C. However, only for some of them (e.g. GRB\,211227A) are the limits sufficiently deep to rule out the full luminosity range observed among stripped-envelope SNe. For some of the remaining bursts, dust extinction could plausibly explain the lack of a detected SN, whereas in other cases the optical coverage was too sparse or shallow to provide meaningful constraints.

As expected, the limits on the KN emission are less constraining.   They rule out a bright KN only for bursts tentatively associated 
with $z\lesssim0.05$ galaxies. 
Interestingly, the available limits also disfavor an AT2017gfo-like optical emission 
in GRB\,211227A which, based on its prompt emission and environment,  is a strong candidate for a merger origin. 
\end{enumerate}

Based on these considerations, we conclude that extrinsic factors, such as 
dust extinction, shallow optical limits, or spurious host associations,  are not sufficient to account for the entire sample of SN-less LGRBs. 
\modify{We identify 15 events, comprising 5 GRBs from Group B and 10 GRBs from Group C with $\mathrm{SFR} \lesssim 1~M_\odot~\mathrm{yr}^{-1}$ and specific SFR  \( \lesssim  3 \,M_\odot\,{L_B^*}^{-1}\,\mathrm{yr}^{-1}\) (Fig.~\ref{fig:sfr}), 
whose properties (prompt emission, host galaxy, and/or lack of SN) are likely explained by a non-collapsar origin. }
The diversity of this sample may extend beyond
the canonical collapsar–merger dichotomy: some events, such as
GRB~051109B and GRB~111005A, do not fit cleanly within either of these
standard progenitor channels.

\begin{figure}
    \centering
    \includegraphics[width=\linewidth]{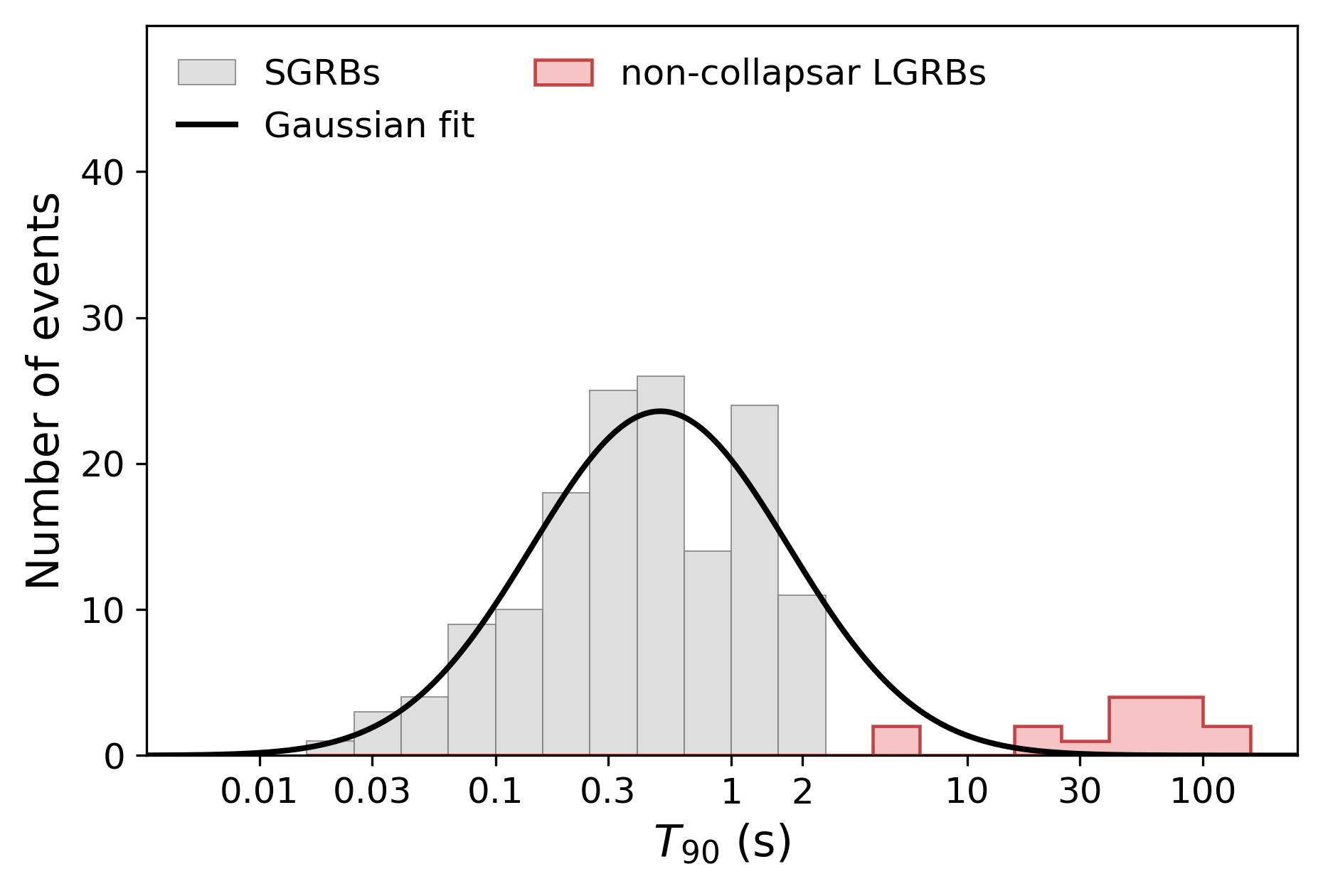}
    \caption{Duration distribution of \textit{Swift} SGRBs, defined by $T_{90}<2$~s. The black curve represents a Gaussian fit to the distribution, centered at 0.5 s with $\sigma$=0.54. 
    The red histogram shows the candidate non-collapsar LGRBs identified at $z\lesssim$0.35.}
    \label{fig:t90}
\end{figure}

The existence of some non-collapsar LGRBs is expected based on the
broad duration distribution of SGRBs  extending beyond the conventional
2~s boundary, just as some SGRBs may arise from the short duration tail of massive star explosions. 
However, we show in Fig.~\ref{fig:t90} that the selected sample
is unlikely to represent a mere extension of the SGRB distribution.
With the exception of GRB\,060505 ($T_{90}\sim$4 s) and GRB\,060912A ($T_{90}\sim$5 s), the rest of the sample
displays substantially longer durations than expected from an
extrapolation of the SGRB $T_{90}$ distribution. 

\subsection{Rate of events}
\label{sec:rate}

We have identified a volume-limited sample of SN-less LGRBs that are unlikely to be produced by a standard collapsar explosion (Sect.~\ref{sec:LGRBs}). 
We derive their volumetric rate as: 

\begin{equation}
R \approx  \frac{4\,\pi}{\Omega (1 - {\rm sin} |b|)} 
\frac{1}{\epsilon\,T}
\frac{1}{f_b}
\eta_c
\sum_{i=1}^N
\frac{1}{V_i}
\end{equation}
where $\Omega$\,$\approx$\,2.2 sr is the BAT FoV, 
$T\simeq20$ yr is the mission lifetime, 
$\epsilon\simeq0.78$ is the average BAT duty cycle, 
$f_b$ is the typical beaming factor, and the factor \(\sin b\) corrects the sky coverage for observational losses along the Galactic plane band \(|b|<10^\circ\). 
The effective volume $V_i$ encodes the burst detectability within the redshift range $z<0.35$, as described in Sect.~\ref{sec:detect}. 
\modify{The factor $\eta_c$ corrects for any residual incompleteness of the sample. Based on our simulations in Sect.~\ref{sec:DisScale}, we find no evidence that a large population of bursts is missing from our sample at $z<0.35$ and estimate $\eta_c \sim 1$.  }

\modify{To define a sample of bursts observed under homogeneous conditions, we exclude from the rate calculation bursts that were not detected and localized through the standard BAT--XRT observing sequence. These are GRB\,060505, a non-slew trigger recovered through manual ground-based analysis, and GRB\,111005A, which lacks an X-ray position. 
To minimize the risk of spurious redshift associations, we exclude bursts with a galactocentric offset larger than 10 arcsec (GRB\,051109B, 060912A, 220611A and 200729A in Fig.~\ref{fig:sfr}). }

\modify{Our most conservative selection (Gold sample) includes those bursts displaying the strongest  evidence for a non-collapsar progenitor, such as GRBs\,050724, 060614, 191019A, and 211211A from Group B.}
The resulting rate of events is $R_{\rm NC} = (0.24 \pm 0.12 )\,f_b^{-1}\ {\rm Gpc^{-3}\,yr^{-1}}$, where the 68\% statistical uncertainty was derived as the variance of the weights $\sigma \propto$ $\sum_i V_i^{-2}$, assuming independent Poissonian events. 

\modify{Our Silver sample comprises new candidates from Group C: GRBs\,050219, 061021, 060428B, 200716C, and 211227A.}  Although the evidence is less compelling for these bursts, 
their global properties support a non-collapsar origin. 

\modify{Our fiducial sample of non-collapsar LGRBs combines the Gold and Silver samples (see Table~\ref{tab:selection} for a summary).} The resulting rate of events is $R_{\rm NC} = (0.5 \pm 0.2 )\,f_b^{-1}\ {\rm Gpc^{-3}\,yr^{-1}}$. 
This estimate is fully consistent with the range
\(0.04\,f_b^{-1}\lesssim R\lesssim0.8\,f_b^{-1}\ {\rm Gpc^{-3}\,yr^{-1}}\)
reported by \citet{Troja2022} and favors its upper end. 
As this paper was nearing completion, an independent study reported a consistent rate estimate \citep{Levan2026}.
No individual burst dominates our inferred rate because the effective volumes, and hence the weights \(V_i^{-1}\), are similar across the sample.

\begin{figure}
    \centering
    \includegraphics[width=\linewidth]{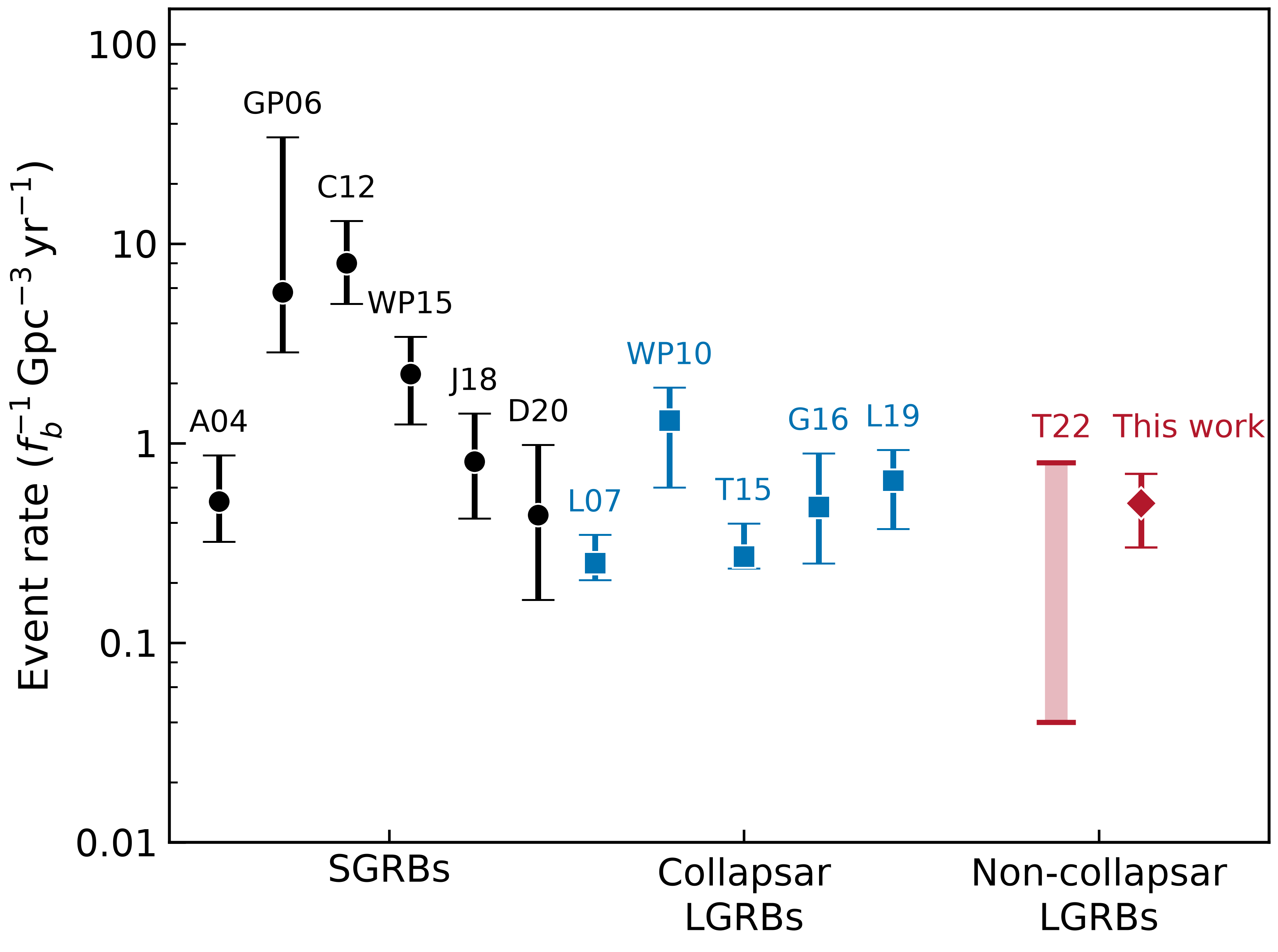}
    \caption{Event rate estimates for SGRBs (black circles), collapsar LGRBs (blue squares), and non-collapsar LGRBs (red diamond),
    not corrected for the unknown beaming factor \(f_b\). 
    All rates are renormalized to a common minimum peak luminosity of \(L_{\min}\approx10^{50}\,{\rm erg\,s^{-1}}\).
    Literature values are taken from: \citealt{Ando04} (A04);  
    \citealt{Guetta06} (GP06); \citealt{Coward12} (C12);
    \citealt{Wanderman15} (WP15); \citealt{Jin2018} (J18);
    \citealt{Dichiara2020} (D20);  \citealt{Liang07} (L07);
    \citealt{Wanderman10} (WP10); \citealt{Tan15} (T15);
    \citealt{Graff16} (G16); \citealt{Lan19} (L19);
    \citealt{Troja2022} (T22). 
    }
    \label{fig:grbrate}
\end{figure}

\begin{figure*}
    \centering
    \includegraphics[width=\linewidth]{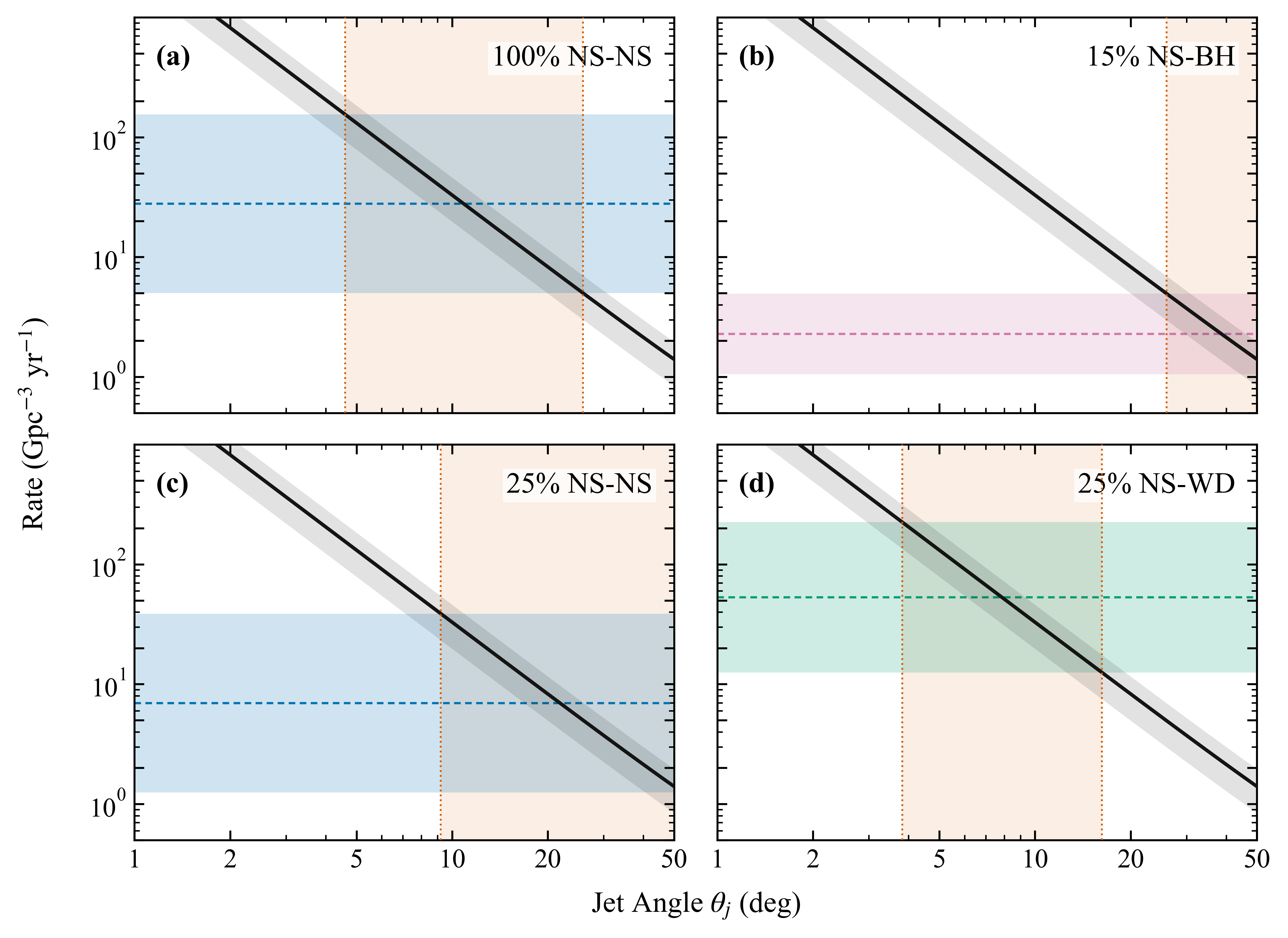}
    \caption{Comparison between the beaming-corrected rate of non-collapsars LGRBs (black curve with 68\% uncertainty shown by the grey band) and the volumetric rates for different compact binary mergers (horizontal shaded bands): (a) NS--NS mergers \citep{GWTC4}, (b) NS--BH mergers \citep{GWTC4}, assuming only $\lesssim$15\% produces an electromagnetic signal \citep{Biscovenu23}, (c) NS--NS mergers, assuming only 25\% produce LGRBs, and (d) NS--WD mergers \citep{He2024}, assuming only 25\% produce LGRBs. 
    The beaming factor is defined as \(f_b=1-\cos\theta_j\), where $\theta_j$ is the jet angle. 
    The vertical bands show the range of jet angles for which the LGRB rate is consistent with the progenitor rate.}
    \label{fig:rates}
\end{figure*}

The three main classes considered here (SGRBs, collapsar LGRBs from Group A, and non-collapsar LGRBs from the Silver sample) contain  comparable numbers of events: 11, 10, and 9, respectively.
This suggests that all three channels contribute to shaping the observed population of nearby high-energy transients. 
In Fig.~\ref{fig:grbrate}, we compare the inferred volumetric rate of non-collapsar LGRBs with 
the rates of SGRBs and collapsar LGRBs from the literature. 
To place these different estimates on a common scale, we homogenize the literature values by renormalizing them to a minimum peak luminosity of $L_{min}\approx10^{50}$\,erg\,s$^{-1}$ using the relevant luminosity function. 

For the population of collapsar LGRBs, the median rate is $R_C \approx0.5 f_{b,\rm C}^{-1}\, {\rm Gpc^{-3}\,yr^{-1}}$, 
comparable to our estimate of non-collapsar LGRBs, conditional on the relative beaming factors of the two populations.
This finding is not consistent with the results of \citet{dong2025statistical}, who determined that SN-less LGRBs
are subdominant at low redshifts. We ascribe this discrepancy 
to their smaller sample (only 8 events) of SN-less events with $z\lesssim$0.35.

A large scatter is visible in the estimated SGRB rates, and we take the values from \citet{Wanderman15} as representative. For 
\(
R_{\rm SGRB}
=
2.2^{+1.2}_{-1.0}\,
f_{b,\rm S}^{-1}\,
{\rm Gpc^{-3}\,yr^{-1}}
\), the ratio between the two populations is therefore:
\[
\frac{R_{\rm NC}}{R_{\rm SGRB}}
=
0.23^{+0.47}_{-0.14}
\left(
\frac{f_{b,\rm SGRB}}{f_{b,\rm NC}}
\right).
\]
Assuming similar beaming corrections ($f_{b,\rm NC} \approx f_{b,\rm SGRB}$),
non-collapsar LGRBs occur at approximately \(23\%\) of the SGRB rate,
with a 68\% credible range between 10\% and 70\%. 
However, as noted in Sect.~\ref{sec:DisScale}, the incidence of chance alignments could be significant in the SGRB sample, potentially causing to overestimate the local event rate. 
A robust determination of the relative contribution therefore requires a careful reassessment of the local SGRB rate (e.g. O'Connor et al., in preparation).

\subsubsection{Comparison to compact binary mergers}

The prime progenitor candidates for the class of non-collapsar LGRBs are compact binary mergers, as indicated by the identification of kilonovae, their host-galaxy properties, and their local environment.
To test whether these progenitor systems could account for all or only some LGRBs, 
in Fig.~\ref{fig:rates} we compare the event rate of non-collapsar LGRBs with the rate of compact binary mergers \citep{GWTC4,Toonen2018,He2024}.  
The shaded horizontal bands indicate the expected volumetric rates for different progenitor channels, whereas the black curve shows the beaming corrected LGRB rate using $f_b=1-\cos\theta_j$ for a two-sided jet.  
We define $\theta_j$ as the maximum angle over which the event would be detected. For a uniform jet, this angle mostly coincides with the jet core half-opening angle ($\theta_j$\,$\approx$\,$\theta_c$), as the flux drops quickly for larger viewing angles. For structured jets $\theta_j$ could be larger than $\theta_c$ and should be interpreted as an effective detection angle. 

The observed LGRB rate is compatible with the NS--NS merger rate for  $\theta_j \simeq 5^\circ$--$25^\circ$ under the assumption that \textit{all}  NS mergers produce a successful jet and a LGRB (panel $a$). 
If only 25\% of NS--NS mergers produce this class of events, the allowed jet angles shift to wider values, $\theta_j \simeq 9^\circ$--$53^\circ$ (panel $c$). 
This comparison shows that, whereas a NS--NS progenitor is possible, the
allowed parameter space is already quite tight: GRB afterglows provide evidence
for narrowly collimated outflows with typical inferred half-opening angles $\theta_c\,\approx5^\circ -15^\circ$ \citep{Burrows2006, Troja2016jet, Jin2018, OConnor2020tale, RoucoEscorial}, which are compatible with the allowed range in panel $a$, but overlap only marginally with the range in panel $c$.

For the same reason, the inferred rate challenges a NS--BH merger progenitor (panel $b$). 
Assuming that $\lesssim$15\% of NS--BH mergers 
produce an electromagnetic (EM) counterpart \citep{Biscovenu23}, an agreement is found only for wide angles $\theta_j \gtrsim25^\circ$ \citep[see also][]{Fishbach2026,Kunnumkai2026}. 
If we instead adopt our most conservative rate estimate from the Gold sample, the tension is alleviated, but a NS--BH origin still requires either a higher EM-bright fraction or prompt emission visible over a broad solid angle ($\theta_j \gtrsim12^\circ$).

Finally, the predicted NS--WD rate shows good agreement with the observations for narrower angles, $\theta_j \simeq 4$--$16^\circ$, even when the fraction of NS--WD mergers producing LGRB jets is modest (panel $d$). 
Unlike NS--NS and NS--BH mergers, whose rates are informed by GW observations, the NS--WD merger rate remains largely based on population-synthesis calculations and indirect empirical estimates \citep[e.g.][]{Toonen2018,He2024}, and therefore carries orders-of-magnitude uncertainty. For these reasons, the NS--WD channel cannot be favored over the other progenitor systems on rate arguments alone. Rather, our comparison shows that this channel represents a viable progenitor for SN-less LGRBs \citep{King2007,Zhong2023}, if they can power a GRB jet \citep{Zenati2019}.

Our conclusions drastically change if we include
the low-luminosity bursts located at low redshift ($z<0.1$) and large offsets ($>$10 arcsec). 
These low-$z$ bursts yield the strongest constraints on the kilonova emission, and, as we show below, their event rate is difficult to reconcile with any merger progenitor. 

Adding GRB\,060912A ($L_{\rm pk, iso}\approx10^{49}$\,erg\,s$^{-1}$) would produce a marginal increase in the event rate, well within its statistical error.  
However, once GRB\,051109B ($L_{\rm pk, iso}\approx6\times10^{47}$\,erg\,s$^{-1}$) is included, the event rate bumps up to $R_{\rm NC} \approx 2 f_b^{-1}\ {\rm Gpc^{-3}\,yr^{-1}}$.
Increasing the observed rate by a factor of 4 increases the requirement on the jet angle by about a factor of 2.
GRBs\,220611A ($L_{\rm pk, iso}\approx10^{47}$\,erg\,s$^{-1}$) would drive the rate up to 
$R_{\rm NC} \approx 4 f_b^{-1}\ {\rm Gpc^{-3}\,yr^{-1}}$, nearly a factor of ten higher than our conservative estimate. 
Therefore, the inclusion of even one of these LGRBs with $L_{\rm pk, iso}\ll 10^{49}$\,erg\,s$^{-1}$ into the sample would be challenging for any merger progenitor model unless their prompt emission is visible over a large solid angle. 
Since none of their afterglow light curves resemble the behavior of an off-axis jet \citep[e.g.][]{Ryan2020}, 
their low luminosities would be the result of very wide jets. 

If instead their emission is collimated into narrow jets, then a different progenitor with a substantially higher rate should be considered. This conclusion is also supported by the tight limits on the associated kilonova. 

GRB~200729A represents an even more extreme case. If associated with its
putative low-redshift host, its low luminosity ($L_{\rm pk, iso}\approx10^{46}$\,erg\,s$^{-1}$) 
would imply a three orders of magnitude increase
in the event rate, $R_{\rm NC} \approx 500 f_b^{-1}\ {\rm Gpc^{-3}\,yr^{-1}}$,
thus exceeding any merger rate.  
Although the simplest solution is that this and the other low-luminosity bursts are spurious redshift associations, it is worth noting that GRB\,200729A would not fit within the Amati relation for any redshift. The prompt emission of this burst suggests a peculiar nature, regardless of its distance scale.

\section{Summary}\label{sec:summary}

We have performed a systematic volume-limited ($z\lesssim$0.35) search for long GRBs not accompanied by bright supernovae using two decades of \textit{Swift} observations. 
The resulting sample consists of 25 events  (Group B and Group C in Tab.~1\ref{tab:sample}) with heterogeneous properties in terms of prompt emission, host galaxies, and local environment. 
By combining these diagnostics, four of them are ruled out as likely chance alignments with a low-redshift galaxy, six more do not lead to a conclusive classification, 
whereas the remaining 15 events appear best explained by a non-collapsar progenitor.

This population is not simply the long duration tail of the short GRB distribution, but constitutes a separate group.  The diversity of its prompt and environmental properties suggests that it may contain multiple progenitor channels rather than representing a single homogeneous class. This possibility is also supported by our inferred event rate.

For our fiducial sample of bursts (Silver sample), we derived an apparent event rate of $\approx 0.5\, {\rm Gpc^{-3}\,yr^{-1}}$ for a standard GRB luminosity. Assuming similar beaming factors, the fiducial rate is comparable to that of LGRBs associated with SNe, and it is consistent with several compact binary merger progenitors. However, attributing the entire population to a single progenitor channel may be challenging. 

At lower redshifts ($z<0.1$), the population is dominated by low-luminosity ($\lesssim$10$^{49}$\,erg\,s$^{-1}$) events, not accompanied by any bright supernova or kilonova emission. These events quickly drive the event rate to higher values, which are difficult to reconcile with standard merger progenitors. 
Owing to the uncertainty in their distance scale, a larger sample with secure redshift measurements would be required to infer their origin.

A targeted search at low redshift appears well suited to this purpose, as it ensures good sensitivity to host galaxies and associated supernova or kilonova emission, allowing meaningful constraints with a modest investment of telescope time. Moreover, we determined that the local balance between the different classes (SGRBs, LGRBs, and SN-less LGRBs) allows us to probe them at an approximately equal rate. 
However, our search has returned fewer than 50 nearby bursts in two decades of observations, which naturally explains the slow progress in the identification of kilonovae and SN-less LGRBs.

One way to increase the sample size would be to extend the search to larger volumes, but this comes at a cost: higher redshifts require more sensitive, highly oversubscribed telescopes, and since the collapsar rate follows the cosmic star formation history, the sample could quickly become dominated by ordinary LGRBs and their SNe.
Therefore, the low-redshift strategy remains attractive because it maximizes diagnostic power, but significant progress would require a more efficient discovery machine. 

An all-sky monitor such as \textit{Fermi}/GBM has likely collected a sample of nearby bursts three times larger than ours, but their coarse localizations limit an efficient identification. 
To combine the benefits of \textit{Swift} and \textit{Fermi}, an all-sky high duty-cycle hard X-ray monitor capable of arcminute localization would be necessary to boost the discovery rate of peculiar transients in the nearby universe.

\begin{acknowledgments}
This work is supported by the European Research Council through the Consolidator grant BHianca (grant agreement ID~101002761). 

%Gemini GRB211227A
Based on observations obtained at the international Gemini Observatory, a program of NSF NOIRLab, which is managed by the Association of Universities for Research in Astronomy (AURA) under a cooperative agreement with the U.S. National Science Foundation on behalf of the Gemini Observatory partnership: the U.S. National Science Foundation (United States), National Research Council (Canada), Agencia Nacional de Investigaci\'{o}n y Desarrollo (Chile), Ministerio de Ciencia, Tecnolog\'{i}a e Innovaci\'{o}n (Argentina), Minist\'{e}rio da Ci\^{e}ncia, Tecnologia, Inova\c{c}\~{o}es e Comunica\c{c}\~{o}es (Brazil), and Korea Astronomy and Space Science Institute (Republic of Korea).

%Keck GRB111225A
Some of the data presented herein were obtained at Keck Observatory, which is a private 501(c)3 non-profit organization operated as a scientific partnership among the California Institute of Technology, the University of California, and the National Aeronautics and Space Administration. The Observatory was made possible by the generous financial support of the W. M. Keck Foundation. The raw Keck Observatory data presented herein, once public, can be accessed from the Keck Observatory Archive (KOA), which is operated by the W. M. Keck Observatory and the NASA Exoplanet Science Institute (NExScI), under contract with the National Aeronautics and Space Administration.

% ESO：GRB150727A X-shooter 095.A-0045(A)，
Based on observations collected at the European Organisation for Astronomical Research in the Southern Hemisphere under ESO programmes 095.A-0045(A), obtained from the ESO Science Archive Facility.

% GTC：GRB111225A 
Based on data from the GTC Archive at CAB (CSIC -INTA). The GTC Archive is part of the Spanish Virtual Observatory project funded by MCIN/AEI/10.13039/501100011033 through grant PID2023-146210NB-I00.

% Subaru GRB111225A
Based in part on data collected at Subaru Telescope and obtained from the SMOKA, which is operated by the Astronomy Data Center, National Astronomical Observatory of Japan.

%SDSS
Funding for the Sloan Digital Sky Survey IV has been provided by the Alfred P. Sloan Foundation, the U.S. Department of Energy Office of Science, and the Participating Institutions. 
SDSS-IV acknowledges support and resources from the Center for High Performance Computing  at the University of Utah. The SDSS website is www.sdss4.org. SDSS-IV is managed by the Astrophysical Research Consortium for the Participating Institutions of the SDSS Collaboration including the Brazilian Participation Group, the Carnegie Institution for Science, Carnegie Mellon University, Center for Astrophysics | Harvard \& Smithsonian, the Chilean Participation Group, the French Participation Group, 
Instituto de Astrof\'isica de Canarias, The Johns Hopkins University, Kavli Institute for the Physics and Mathematics of the Universe (IPMU) / University of Tokyo, the Korean Participation Group, Lawrence Berkeley National Laboratory, Leibniz Institut f\"ur Astrophysik Potsdam (AIP),  Max-Planck-Institut f\"ur Astronomie (MPIA Heidelberg), Max-Planck-Institut f\"ur Astrophysik (MPA Garching), Max-Planck-Institut f\"ur Extraterrestrische Physik (MPE), National Astronomical Observatories of China, New Mexico State University, New York University, University of Notre Dame, Observat\'orio Nacional / MCTI, The Ohio State University, Pennsylvania State University, Shanghai Astronomical Observatory, United Kingdom Participation Group, Universidad Nacional Aut\'onoma de M\'exico, University of Arizona, University of Colorado Boulder, University of Oxford, University of Portsmouth, University of Utah, University of Virginia, University of Washington, University of Wisconsin, Vanderbilt University, and Yale University.

%Legacy survey
The Legacy Surveys consist of three individual and complementary projects: the Dark Energy Camera Legacy Survey (DECaLS; Proposal ID \#2014B-0404; PIs: David Schlegel and Arjun Dey), the Beijing-Arizona Sky Survey (BASS; NOAO Prop. ID \#2015A-0801; PIs: Zhou Xu and Xiaohui Fan), and the Mayall z-band Legacy Survey (MzLS; Prop. ID \#2016A-0453; PI: Arjun Dey). DECaLS, BASS and MzLS together include data obtained, respectively, at the Blanco telescope, Cerro Tololo Inter-American Observatory, NSF's NOIRLab; the Bok telescope, Steward Observatory, University of Arizona; and the Mayall telescope, Kitt Peak National Observatory, NOIRLab. Pipeline processing and analyses of the data were supported by NOIRLab and the Lawrence Berkeley National Laboratory (LBNL). The Legacy Surveys project is honored to be permitted to conduct astronomical research on I'oligam Du'ag (Kitt Peak), a mountain with particular significance to the Tohono O'odham Nation.
NOIRLab is operated by the Association of Universities for Research in Astronomy (AURA) under a cooperative agreement with the National Science Foundation. LBNL is managed by the Regents of the University of California under contract to the U.S. Department of Energy.
This project used data obtained with the Dark Energy Camera (DECam), which was constructed by the Dark Energy Survey (DES) collaboration. Funding for the DES Projects has been provided by the U.S. Department of Energy, the U.S. National Science Foundation, the Ministry of Science and Education of Spain, the Science and Technology Facilities Council of the United Kingdom, the Higher Education Funding Council for England, the National Center for Supercomputing Applications at the University of Illinois at Urbana-Champaign, the Kavli Institute of Cosmological Physics at the University of Chicago, Center for Cosmology and Astro-Particle Physics at the Ohio State University, the Mitchell Institute for Fundamental Physics and Astronomy at Texas A\&M University, Financiadora de Estudos e Projetos, Fundacao Carlos Chagas Filho de Amparo, Financiadora de Estudos e Projetos, Fundacao Carlos Chagas Filho de Amparo a Pesquisa do Estado do Rio de Janeiro, Conselho Nacional de Desenvolvimento Cientifico e Tecnologico and the Ministerio da Ciencia, Tecnologia e Inovacao, the Deutsche Forschungsgemeinschaft and the Collaborating Institutions in the Dark Energy Survey. The Collaborating Institutions are Argonne National Laboratory, the University of California at Santa Cruz, the University of Cambridge, Centro de Investigaciones Energeticas, Medioambientales y Tecnologicas-Madrid, the University of Chicago, University College London, the DES-Brazil Consortium, the University of Edinburgh, the Eidgenossische Technische Hochschule (ETH) Zurich, Fermi National Accelerator Laboratory, the University of Illinois at Urbana-Champaign, the Institut de Ciencies de l’Espai (IEEC/CSIC), the Institut de Fisica d’Altes Energies, Lawrence Berkeley National Laboratory, the Ludwig Maximilians Universitat Munchen and the associated Excellence Cluster Universe, the University of Michigan, NSF's NOIRLab, the University of Nottingham, the Ohio State University, the University of Pennsylvania, the University of Portsmouth, SLAC National Accelerator Laboratory, Stanford University, the University of Sussex, and Texas A\&M University.
BASS is a key project of the Telescope Access Program (TAP), which has been funded by the National Astronomical Observatories of China, the Chinese Academy of Sciences (the Strategic Priority Research Program ``The Emergence of Cosmological Structures" Grant \# XDB09000000), and the Special Fund for Astronomy from the Ministry of Finance. The BASS is also supported by the External Cooperation Program of Chinese Academy of Sciences (Grant \# 114A11KYSB20160057), and Chinese National Natural Science Foundation (Grant \# 12120101003, \# 11433005).
The Legacy Survey team makes use of data products from the Near-Earth Object Wide-field Infrared Survey Explorer (NEOWISE), which is a project of the Jet Propulsion Laboratory/California Institute of Technology. NEOWISE is funded by the National Aeronautics and Space Administration.
The Legacy Surveys imaging of the DESI footprint is supported by the Director, Office of Science, Office of High Energy Physics of the U.S. Department of Energy under Contract No. DE-AC02-05CH11231, by the National Energy Research Scientific Computing Center, a DOE Office of Science User Facility under the same contract; and by the U.S. National Science Foundation, Division of Astronomical Sciences under Contract No. AST-0950945 to NOAO.

\end{acknowledgments}

\appendix
\renewcommand{\thetable}{A\arabic{table}}
\renewcommand{\thefigure}{A\arabic{figure}}
\setcounter{table}{0}
\setcounter{figure}{0}

\section{Host galaxy properties}\label{app:host}
\begin{figure}
    \centering
    \includegraphics[width=0.98\linewidth]{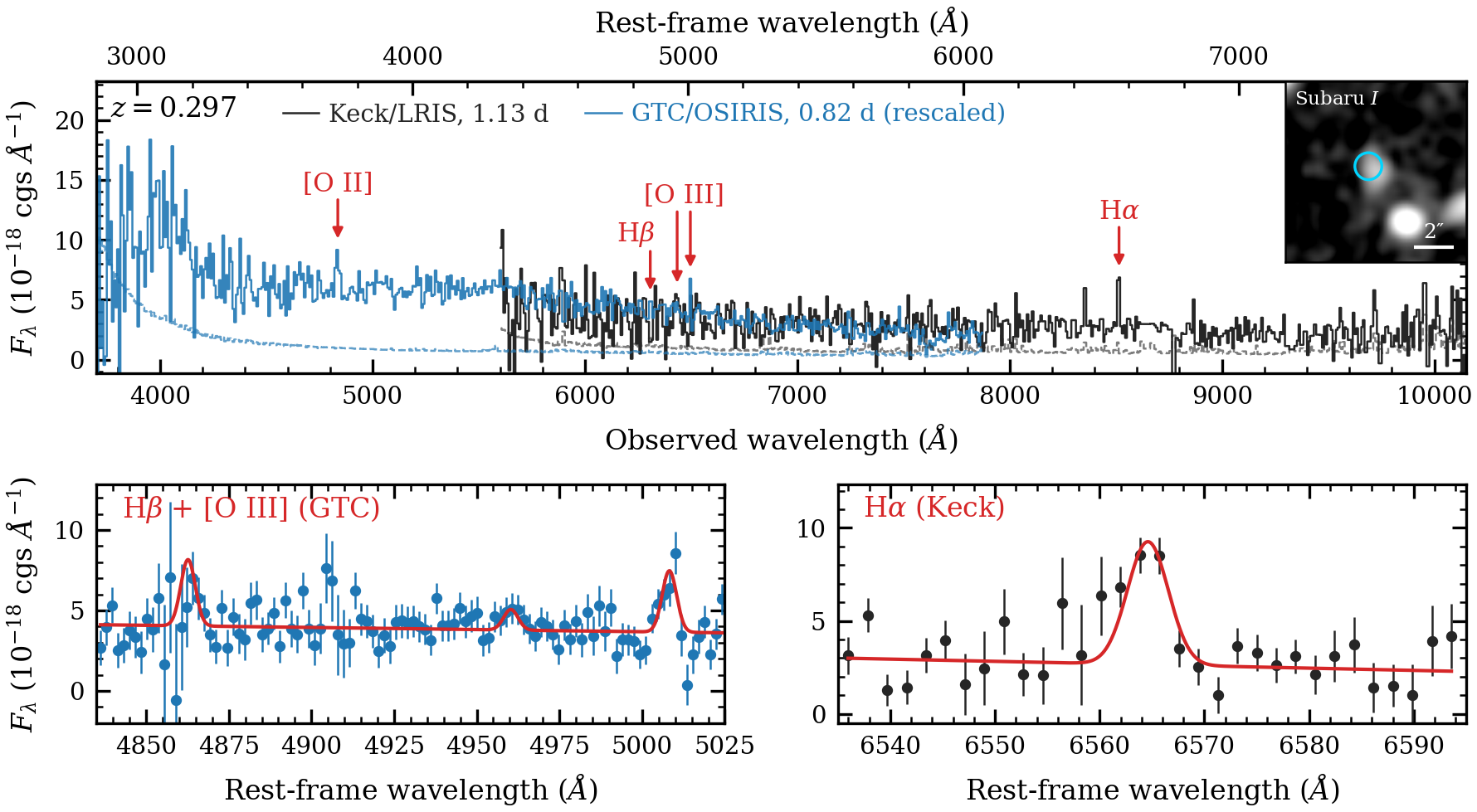}
    \caption{\modify{Keck/LRIS ($T_0+0.82$\,d, black) and GTC/OSIRIS \citep[$T_0+1.13$\,d, blue;][]{Thone2014GCN16079} spectra of GRB\,111225A. The Keck spectrum is scaled to the same-night LRIS $R$-band photometry to correct for slit losses, whereas the GTC spectrum is rescaled for plotting purposes. Dashed lines are the corresponding error spectra. The nebular lines are marked. 
    The inset shows the GRB localization (circle) on top of its faint host galaxy. 
    The lower panels show our modeling of the $H\alpha$ and $H\beta$ lines. H$\beta$ is strongly affected by sky emission and only marginally detected in the GTC spectrum. 
    }}
    \label{fig:111225Ahostspec}
\end{figure}

\begin{figure}
    \centering
    \includegraphics[width=0.98\linewidth]{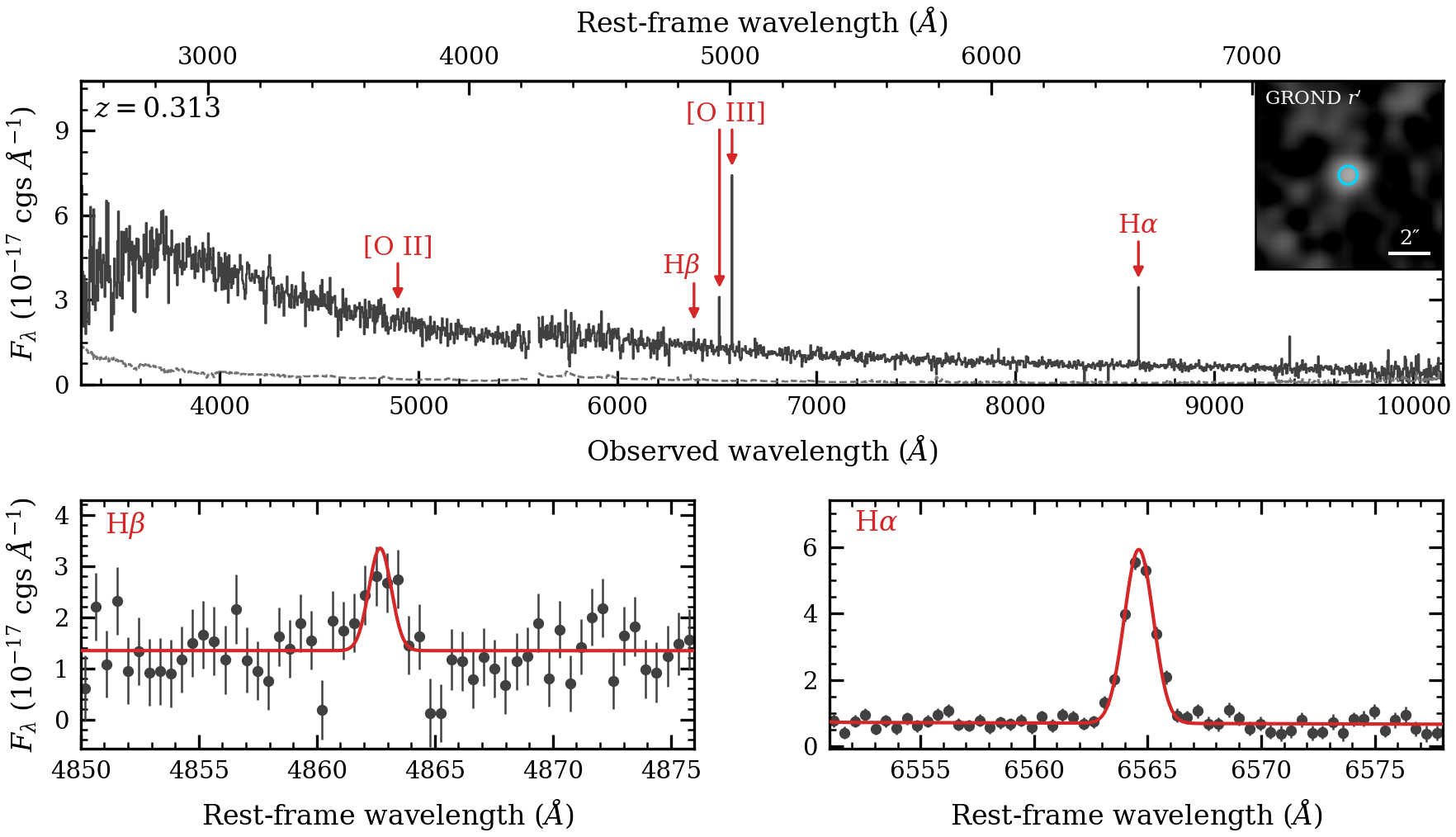}
    \caption{\modify{VLT/X-shooter spectrum (black solid curve) and corresponding error spectrum (black dashed curve) of GRB 150727A taken at $T_0+0.23$\,d \citep{Selsing2019}. The brightest nebular lines are marked. The inset shows the afterglow localization (circle) on top of its faint galaxy.    
    The lower panels show our modeling of the $H\alpha$ and $H\beta$ lines.}  }
    \label{fig:150727Ahostspec}
\end{figure}

\modify{Tab.~\ref{tab:sfr} lists the star formation rates and rest-frame $B$-band luminosities of the host galaxies shown in Fig.~\ref{fig:sfr}. SFRs are taken from the literature when available. 
In three cases (GRB\,111225A,  GRB\,150727A, and GRB\,200716C) we analyzed the public spectra
to derive the SFR from emission lines. 
We modelled the continuum locally and fitted the emission lines as Gaussians. Line fluxes were obtained by integrating the best-fit Gaussian profiles, and their uncertainties were estimated using 1000 Monte Carlo simulations.
SFRs were derived from the H$\alpha$ luminosity using the calibration of \citet{Kennicutt2012}, $\mathrm{SFR}=5.37\times10^{-42}\,L(\mathrm{H}\alpha)$\,M$_\odot$\,yr$^{-1}$ with a Chabrier initial mass function \citep{Chabrier2003}, or, when H$\alpha$ falls outside the observed band, from [O\,{\sc ii}]\,$\lambda3727$. 
The spectrum of GRB\,200716C is shown in \citet{Giarratana2023}, whereas the spectra of GRB\,111225 and GRB\,150727A are shown in Fig.~\ref{fig:111225Ahostspec} and Fig.~\ref{fig:150727Ahostspec}.
}

Rest-frame $B$-band absolute magnitudes are inferred from the host photometry closest in rest-frame wavelength to the $B$ band, K-corrected to rest-frame $B$, and normalized to $M_B^* = -20.5$ \citep{Tortorelli2020}.

\section{Gemini Observations of GRB~211227A}
We retrieved archival Gemini-South imaging of GRB\,211227A obtained under programme GS-2021B-DD-105 (PI: B. O'Connor). Observations with the Gemini Multi-Object Spectrograph on Gemini South
(GMOS-S) consisted of  $15\times120$\,s exposures in the $i$ band and  were acquired on 2021 December 31 and 2022 January 9, at a mean times of 3.3 and 12.4 days after the burst, respectively. 
We also analyzed Flamingos-2 (F2) $J$-band observations carried out on January 2  for a total exposure of 180 s. The images were reduced using \texttt{DRAGONS} v3.2.0 \citep{Labrie2019,Labrie2023} and associated calibration frames. Photometric zero points were determined from unsaturated
Pan-STARRS DR2 \citep{Magnier2020} stars for $i$ and 2MASS \citep{2mass} stars for $J$. 
Within the XRT position, we found no optical counterpart down to $i>26.2$ AB mag and $i>25.7$ AB mag at 3.3\,d and 12.4\,d post GRB trigger. We place an upper limits of $J>22.5$ AB mag at 5.3 d. 
All magnitudes are corrected for galactic extinction in this direction ($A_V\approx$0.05 mag). 
At $z=0.228$ the late $i$-band limit places a deep constraint on the associated SN (Tab. \ref{tab:constraints}).

\section{Methodology of the supernova and kilonova comparison}
\label{app:snkn}

For each GRB in our sample, we assess whether the available dataset is compatible with an accompanying transient with the luminosity and evolution of SN~1998bw or AT2017gfo. 
For each burst, we adopt the redshift of the associated galaxy and
convert the observer's frame time and the effective wavelength of the observing filter into the rest-frame. The observed photometry is corrected for extinction and distance modulus, then compared with the template light curve evaluated at the same rest-frame time and wavelength. 

For each measurement, we evaluate the quantity $\Delta M = M_{\rm obs} - M_{\rm template}$ across all available epochs and filters, and record its most constraining value in Tab.~1\ref{tab:constraints}. 
With this sign convention, $\Delta M>0$ means that the observed measurement lies below the template prediction, and any accompanying transient must therefore be fainter than the template by at least
$\Delta M$. Vice versa, $\Delta M\leq0$ means that the observation alone does not exclude the template brightness.
We do not make any assessment on the nature of the observed emission, and whether it is afterglow, kilonova, or supernova-dominated.  A non-detection of the GRB counterpart places an upper limit on the flux of any accompanying transient. A detection also provides a flux ceiling, as any supernova or kilonova contribution cannot exceed the total measured flux. 

For SN\,1998bw, we use the de-reddened and K-corrected \textit{UBVRI} light curves of \citet{Clocchiatti2011}, covering rest-frame epochs from 1.9 to 498\,d, complemented by the sparse NIR photometry of \citet{Patat2001} and \citet{Sollerman2002}. At each epoch, the spectrum \texttt{v19-1998bw} \citep{Vincenzi2019} is rescaled to reproduce the available photometry. For AT2017gfo, the reference light curve is compiled from the literature. The original sources of data are \citet{Andreoni2017}, \citet{Arcavi2017},
\citet{Coulter2017}, \citet{Cowperthwaite2017}, \citet{Diaz2017}, \citet{Drout2017}, \citet{Pian2017}, \citet{Smartt2017}, \citet{Troja2017}, \citet{Tanvir2017}, \citet{Utsumi2017}.

The GRB photometry was compiled from the literature, and individual sources are cited in Tab.~1\ref{tab:constraints}. 
The \texttt{grbLC}\footnote{\url{https://github.com/SLAC-Gamma-Rays/grbLC}} optical GRB afterglow catalogue \citep{Dainotti2024} was used to identify the original literature references for part of the photometric data.
\textit{Swift}/UVOT photometry was taken from the Second \textit{Swift} UVOT GRB Afterglow Catalog\footnote{\url{https://swift.gsfc.nasa.gov/results/uvot_grbcat2/}} \citep{Roming2017} and rebuilt from the tabulated source counts using the zero points of \citet{Breeveld2011}.

\clearpage 

\input{Tab/sfr_luminosity}
\clearpage
\clearpage
\input{Tab/constraints}

\clearpage

\input{Tab/selection}

\bibliography{main}{}
\bibliographystyle{aasjournalv7}

\end{document}

%% file: Tab/sample.tex
\begin{deluxetable*}{lclclcrc}
\ifdefined\rotateonfalse\rotateonfalse\fi
\tablecaption{
Sample of nearby ($z$\,$\lesssim$0.35) GRBs divided into four sub-groups: short duration bursts (SGRBs), long bursts with secure SN (group A), 
SN-less long bursts (group B), and the
remaining sample of long bursts (group C). 
\label{tab:sample}
}
% \tabletypesize{\small}
\tabletypesize{\tiny}
\tablewidth{0pt}
\setlength{\tabcolsep}{15pt} 
\tablehead{
\colhead{GRB} & 
\colhead{$T_{90}$ [s]} &
\colhead{$z$} &
\colhead{SN} & 
\colhead{$P_{\rm cc}$}&
\colhead{$A_{V,\rm Gal}$}&
\colhead{$A_{V,z}$}&
\colhead{Ref.}
}
\startdata
\tableline
%%%%%%%%%%%%%%%%%%%%%%%%%%%%%%%%%%%%%%%%%%%%%%%%%%%%%%%%%
%%%%%%%%%%%%%%%%%%%%%%%%%%%%%%%%%%%%%%%%%%%%%%%%%%%%%%%%%
%%%%%%%%%%%%%%%%%%%%%   SGRBs   %%%%%%%%%%%%%%%%%%%%%%%%%
%%%%%%%%%%%%%%%%%%%%%%%%%%%%%%%%%%%%%%%%%%%%%%%%%%%%%%%%%
%%%%%%%%%%%%%%%%%%%%%%%%%%%%%%%%%%%%%%%%%%%%%%%%%%%%%%%%%
\multicolumn{8}{l}{\textbf{SGRBs}}\\
\hline
050509B  & $0.024\pm0.009$ & 0.225  & N & 0.002  & 0.05 & \nodata  & \hyperlink{Gehrels2005}{1}\\
060502B  & $0.14\pm0.05$ & 0.287  &  N  & 0.03 & 0.12 & \nodata  & \hyperlink{Bloom2007}{2}\\
070809 & $1.3\pm0.4$ & 0.2187 &  N & 0.06  & 0.25& \nodata  & \hyperlink{Jin2020}{3}\\
080905A & $1.02\pm0.08$ & 0.1218  &  N & 0.003 & 0.38& \nodata & \hyperlink{Rowlinson2010}{4}\\
100628A  & $0.036\pm0.009$ & 0.102  &  \nodata & 0.12 & 0.47 & \nodata  &  \hyperlink{NicuesaGuelbenzu2015}{5}\\
120305A & $0.10\pm0.02$ & 0.225  & \nodata & 0.05  & 1.12& \nodata  & \hyperlink{Fong2022}{6}\\
130822A &$0.044\pm0.010$& 0.154  & N & 0.03  & 0.07& \nodata & \hyperlink{Oconnor2022}{7}\\
150101B &$0.012\pm0.009$& 0.134  & N & 0.0003 & 0.11& \nodata 
& \hyperlink{Fong2016}{8},\hyperlink{Troja2018}{9}\\
160821B &$0.48\pm0.07$& 0.161  & N & 0.008  & 0.12& \nodata & \hyperlink{Troja2019}{10} \\
210919A &$0.16\pm0.03$& 0.2415  & N & 0.07  & 0.39& \nodata  
& \hyperlink{Fong2022}{6},\hyperlink{OConnor2021GCN210919A}{11}\\
231117A &$0.67\pm0.07$& 0.257 &  N & 0.001 & 0.20 & \nodata & \hyperlink{Schroeder2025}{12}\\\tableline 
%%%%%%%%%%%%%%%%%%%%%%%%%%%%%%%%%%%%%%%%%%%%%%%%%%%%%%%%%
%%%%%%%%%%%%%%%%%%%%%%%%%%%%%%%%%%%%%%%%%%%%%%%%%%%%%%%%%
%%%%%%%%%%%%%%%%%%%%%  Group A  %%%%%%%%%%%%%%%%%%%%%%%%%
%%%%%%%%%%%%%%%%%%%%%%%%%%%%%%%%%%%%%%%%%%%%%%%%%%%%%%%%%
%%%%%%%%%%%%%%%%%%%%%%%%%%%%%%%%%%%%%%%%%%%%%%%%%%%%%%%%%
\multicolumn{8}{l}{\textbf{Group A}}\\
\hline
060218 &$2100\pm100$& 0.03342  & SN\,2006aj & 0.004 & 0.40 & 0.13
&\hyperlink{Pian2006}{13},\hyperlink{Guenther2006}{14} \\ 
100316D &$500 \pm 400$& 0.0591  & SN\,2010bh & 0.001 & 0.32& 0.43 
& \hyperlink{Starling2011}{15},\hyperlink{Bufano2012}{16}\\ 
120422A & $60 \pm 6$&0.2826  & SN\,2012bz & 0.008 & 0.09& \nodata 
& \hyperlink{Schulze2014}{17}\\ 
130427A & $244 \pm 5$&0.3399  & SN\,2013cq & 0.003 & 0.06& 0.16 & \hyperlink{Xu2013}{18}\\ 
150818A & $140\pm20$&0.282  & SN Ic-BL & 0.007 &0.06& \nodata 
& \hyperlink{SanchezRamirez2015}{19},\hyperlink{GCN18213}{20}\\
161219B & $6.9\pm0.8$&0.1475  & SN\,2016jca& 0.01 & 0.09& 0.05 
& \hyperlink{Cano2017}{21}\\ 
171205A & $190\pm30$&0.0368 & SN\,2017iuk & 0.0003 & 0.14& 0.06 
& \hyperlink{Izzo2019}{22}\\ 
180728A & $8.7\pm0.3$&0.1171 & SN\,2018fip & 0.004 & 0.76& 0.04 
& \hyperlink{Rossi2026}{23}\\ 
190829A & $60\pm50$&0.0785 & SN\,2019oyw & 0.0006 &0.13& 2.8 & 
\hyperlink{Hu2021}{24},\hyperlink{Dichiara2022}{25}\\ 
221009A &$1105\pm8$& 0.1510  & SN\,2022xiw & 0.004 & 4.22 & 0.1 
& \hyperlink{Srinivasaragavan2023}{26}\\
\tableline
%%%%%%%%%%%%%%%%%%%%%%%%%%%%%%%%%%%%%%%%%%%%%%%%%%%%%%%%%
%%%%%%%%%%%%%%%%%%%%%%%%%%%%%%%%%%%%%%%%%%%%%%%%%%%%%%%%%
%%%%%%%%%%%%%%%%%%%%%  Group B  %%%%%%%%%%%%%%%%%%%%%%%%%
%%%%%%%%%%%%%%%%%%%%%%%%%%%%%%%%%%%%%%%%%%%%%%%%%%%%%%%%%
%%%%%%%%%%%%%%%%%%%%%%%%%%%%%%%%%%%%%%%%%%%%%%%%%%%%%%%%%
\multicolumn{8}{l}{\textbf{Group B}}\\
\hline
050724A & $99\pm9$&0.257   & N & 0.0001  & 1.62& \nodata 
& \hyperlink{Barthelmy2005}{27},\hyperlink{Berger2005}{28}\\
060505 &$4.0\pm1.0$& 0.089 & N & 0.0009 & 0.06& \nodata & 
\hyperlink{Fynbo2006}{29},\hyperlink{Ofek2007}{30} \\
060614 & $109\pm3$&0.125 & N & 0.004 & 0.06& 0.05 & 
\hyperlink{Gehrels2006}{31},\hyperlink{Yang2015}{32},\hyperlink{Mangano2007}{33}\\ 
191019A & $64\pm5$&0.248 & N &  0.00008  & 0.10& 0.06 & 
\hyperlink{Levan2023}{34} \\
211211A &$50.7\pm0.9$& 0.076 & N & 0.01 & 0.05& $<$0.02 & 
\hyperlink{Troja2022}{35},\hyperlink{Yang22}{36},\hyperlink{Rastinejad2022}{37}\\ %Avz from Troja2022
\tableline
%%%%%%%%%%%%%%%%%%%%%%%%%%%%%%%%%%%%%%%%%%%%%%%%%%%%%%%%%
%%%%%%%%%%%%%%%%%%%%%%%%%%%%%%%%%%%%%%%%%%%%%%%%%%%%%%%%%
%%%%%%%%%%%%%%%%%%%%%  Group C  %%%%%%%%%%%%%%%%%%%%%%%%%
%%%%%%%%%%%%%%%%%%%%%%%%%%%%%%%%%%%%%%%%%%%%%%%%%%%%%%%%%
%%%%%%%%%%%%%%%%%%%%%%%%%%%%%%%%%%%%%%%%%%%%%%%%%%%%%%%%%
\multicolumn{8}{l}{\textbf{Group C}}\\
\hline
050219A &$24\pm2$& 0.211  & \nodata & 0.006 & 0.45 & \nodata & \hyperlink{Rossi2014}{38}\\ 
050826 &$30\pm6$& 0.296 &  \nodata  &0.004&1.70& $\lesssim$0.6 & \hyperlink{Mirabal2007}{39}\\ 
050911$^{\dagger}$&$16.2 \pm 0.9$&  0.1646 &  \nodata  & 0.003  & 0.03 & \nodata & \hyperlink{Berger2007}{40}\\
051109B &$16\pm4$& 0.080  &  N  & 0.001 &0.45 & 
$\lesssim$1.8 & \hyperlink{Perley2006GCN}{41}\\ 
060428B &$96 \pm 50$ & 0.348 &  N  &0.003&0.04& $\lesssim$0.7 & \hyperlink{Perley2007AIP}{42} \\
060912A &$5.0\pm0.6$ & 0.0939   &  N & 0.0008$^\ddagger$ & 0.14 &  $\approx$\,0.20 & 
\hyperlink{Abdurrouf2022}{43},\hyperlink{Covino2013}{44} \\ 
061021 &$48\pm6$& 0.3463  &  \nodata  &0.03& 0.16 &$<$0.06 & \hyperlink{Covino2013}{44},\hyperlink{Fynbo2009}{45}\\
070412$^{\dagger}$ & $34\pm5$ & 0.0307& \nodata & 0.0002 &  0.06 & $\gtrsim$3 & \hyperlink{Rol2007}{46} \\
070521$^{\dagger}$ &$39\pm2$ & 0.031   &  \nodata  & 0.0007&  0.07& $\gtrsim$2 & \hyperlink{Ofek2007GCN}{47} \\ 
080517 &$65\pm20$& 0.089  &  \nodata & 0.0003 & 0.29 & \nodata & \hyperlink{Stanway2015}{48} \\ 
090417B &$270\pm30$& 0.345  &  \nodata & 0.01 & 0.05&$\gtrsim$12 & \hyperlink{Holland2010}{49}\\ 
111005A &$23\pm5$& 0.01326  &  N &  0.0005& 0.25& $\sim$2 (Host) & \hyperlink{Michalowski2018}{50},\hyperlink{Tanga2018}{51}\\ 
111225A &$110\pm30$& 0.297 &  \nodata  & 0.02& 0.71 & $\lesssim$0.4 & \hyperlink{Thone2014GCN16079}{52}\\ 
130925A &$160\pm3$& 0.347  &  \nodata  & 0.004& 0.06 & $\sim$5 
& \hyperlink{Greiner2014}{53},\hyperlink{Schady2015}{54}\\ 
150424A$^{\dagger}$ & $81\pm17$& 0.30   &  \nodata  & 0.02& 0.16 & $\lesssim$0.4 
& \hyperlink{AlbertoGCN}{55}  \\
150727A &$88\pm11$& 0.313 &  \nodata  &  \modify{0.01} & 0.24 & $\lesssim$2 
& \hyperlink{Selsing2019}{56}\\ 
200716C & $87\pm15$  & 0.341 &  \nodata & 0.001 & 0.04& $\lesssim$0.2 
& \hyperlink{Giarratana2023}{57} \\ 
200729A &$120\pm30$& 0.00176  & N & 0.0002 & 0.03 & \nodata 
& \hyperlink{Krimm2020GCN}{58},\hyperlink{Dichiara2020GCN28270}{59}\\ 
211227A &$84\pm8$& 0.228 & N & 0.005 & 0.05 & $\lesssim$0.7 & \hyperlink{Lu2022}{60}\\
220611A &$57\pm12$ &0.049 & N & 0.0002& 0.09 & $\lesssim$1 & \hyperlink{Cenko2022}{61}\\
\enddata
\tabletypesize{\scriptsize}

\begingroup
\renewcommand{\tablenotetext}[2]{\vskip1pt{\currtabletypesize\indent\vrule height 7pt depth 2pt width 0pt\relax$^{#1}$\,#2\par}}
\tablenotetext{\dagger}{Event classified in Sect.~\ref{sec:environment} as a likely chance alignment with a low-redshift galaxy.}
\tablenotetext{\ddagger}{\modify{This value is lower than the value 
quoted in \citet{Levan2007}. The latter was derived using the galaxy number counts of \citet{Hogg1997}, calibrated for magnitudes fainter than $R\sim20$\,mag.}}
\endgroup

\tablerefs{
%%%%%%%%%%%%%%%%%%%%%%%%%% SGRB %%%%%%%%%%%%%%%%%%%%%%%%%%
%050509B
 \hypertarget{Gehrels2005}{(1)}~\citet{Gehrels2005};
%060502B
\hypertarget{Bloom2007}{(2)}~\citet{Bloom2007};
%070809
 \hypertarget{Jin2020}{(3)}~\citet{Jin2020};
%080905A
\hypertarget{Rowlinson2010}{(4)}~\citet{Rowlinson2010};
%100628A 
\hyperlink{NicuesaGuelbenzu2015}{(5)}~\citet{NicuesaGuelbenzu2015};
%120305A 
\hypertarget{Fong2022}{(6)}~\citet{Fong2022};
%130822A
 \hypertarget{OConnor2022}{(7)}~\citet{OConnor2022};
%150101B
\hypertarget{Fong2016}{(8)}~\citet{Fong2016};
\hypertarget{Troja2018}{(9)}~\citet{Troja2018};
%160821B
\hypertarget{Troja2019}{(10)}~\citet{Troja2019};
%210919A  galaxy
\hypertarget{OConnor2021GCN210919A}{(11)}~\citet{OConnor2021GCN210919A};
%231117A 
\hypertarget{Schroeder2025}{(12)}~\citet{Schroeder2025};
%%%%%%%%%%%%%%%%%%%%%%%%%% Group A %%%%%%%%%%%%%%%%%%%%%%%%%%
%  060218
\hypertarget{Pian2006}{(13)}~\citet{Pian2006}; % SN association
\hypertarget{Guenther2006}{(14)}~\citet{Guenther2006}; %Avz NaID
% 100316D
\hypertarget{Starling2011}{(15)}~\citet{Starling2011}; % z 
\hypertarget{Bufano2012}{(16)}~\citet{Bufano2012}; % Avz NaID & SN
% 120422A
\hypertarget{Schulze2014}{(17)}~\citet{Schulze2014}; % z; SN; negligible dust extinction
%130427A
\hypertarget{Xu2013}{(18)}~\citet{Xu2013}; %z, SN, Avz NaID and griz&X-ray SED 
%150818A
\hypertarget{SanchezRamirez2015}{(19)}~\citet{SanchezRamirez2015}; %z
\hypertarget{GCN18213}{(20)}~\citet{GCN18213}; %SN
%161219B
\hypertarget{Cano2017}{(21)}~\citet{Cano2017}; %SN; AVz from line-of-sight from NIR-to-X-ray afterglow SED
%171205A
\hypertarget{Izzo2019}{(22)}~\citet{Izzo2019}; %z; Avz consistent from NaID and Balmer decrement; SN
%180728A
\hypertarget{Rossi2026}{(24)}~\citet{Rossi2026}; %z, SN, Avz from optical-X-ray SED and better LMC
%190829A
\hypertarget{Hu2021}{(24)}~\citet{Hu2021}; %z, SN 
\hypertarget{Dichiara2022}{(25)}~\citet{Dichiara2022}; %Avz
%221009A 
\hypertarget{Srinivasaragavan2023}{(26)}~\citet{Srinivasaragavan2023}; 
%%%%%%%%%%%%%%%%%%%%%%%%%% Group B %%%%%%%%%%%%%%%%%%%%%%%%%%
%050724A 
\hypertarget{Barthelmy2005}{(27)}~\citet{Barthelmy2005}; 
\hypertarget{Berger2005}{(28)}~\citet{Berger2005}; 
%060505
\hypertarget{Fynbo2006}{(29)}~\citet{Fynbo2006}; 
\hypertarget{Ofek2007}{(30)}~\citet{Ofek2007}; 
%060614
\hypertarget{Gehrels2006}{(31)}~\citet{Gehrels2006}; 
\hypertarget{Yang2015}{(32)}~\citet{Yang2015}; 
\hypertarget{Mangano2007}{(33)}~\citet{Mangano2007}
% 191019A
\hypertarget{Levan2023}{(34)}~\citet{Levan2023}; 
%211211
\hypertarget{Troja2022}{(35)}~\citet{Troja2022}; 
\hypertarget{Yang2022}{(36)}~\citet{Yang2022}; 
\hypertarget{Rastinejad2022}{(37)}~\citet{Rastinejad2022}; 
%%%%%%%%%%%%%%%%%%%%%%%%%% Group C %%%%%%%%%%%%%%%%%%%%%%%%%%
\hypertarget{Rossi2014}{(38)}~\citet{Rossi2014}; %050219A
\hypertarget{Mirabal2007}{(39)}~\citet{Mirabal2007}; %050826
\hypertarget{Berger2007}{(40)}~\citet{Berger2007};%050911
\hypertarget{Perley2006GCN}{(41)}~\citet{Perley2006GCN};%051109B
\hypertarget{Perley2007AIP}{(42)}~\citet{Perley2007AIP}; %060428B
% \hypertarget{Levan2007}{(43)}~\citet{Levan2007}; %060912A another z=0.937 not quoted here
\hypertarget{Abdurrouf2022}{(43)}~\citet{Abdurrouf2022}; %060912A SDSS DR17 z=0.09391±0.00002 
\hypertarget{Covino2013}{(44)}~\citet{Covino2013}; 
\hypertarget{Fynbo2009}{(45)}~\citet{Fynbo2009}; %061021
\hypertarget{Rol2007}{(46)}~\citet{Rol2007}; %070412
\hypertarget{Ofek2007GCN}{(47)}~\citet{Ofek2007GCN}; %070521
\hypertarget{Stanway2015}{(48)}~\citet{Stanway2015}; %080517
\hypertarget{Holland2010}{(49)}~\citet{Holland2010}; %090417B 
\hypertarget{Michalowski2018}{(50)}~\citet{Michalowski2018}; %111005A 
\hypertarget{Tanga2018}{(51)}~\citet{Tanga2018}; %111005A Av
\hypertarget{Thone2014GCN16079}{(52)}~\citet{Thone2014GCN16079}; %111225A 
\hypertarget{Greiner2014}{(53)}~\citet{Greiner2014};  %130925
\hypertarget{Schady2015}{(54)}~\citet{Schady2015}; %130925
\hypertarget{AlbertoGCN}{(55)}~\citet{Castro-Tirado2015}; %150424A
\hypertarget{Selsing2019}{(56)}~\citet{Selsing2019}; %150727A 
\hypertarget{Giarratana2023}{(57)}~\citet{Giarratana2023}; %200716C 
\hypertarget{Krimm2020GCN}{(58)}~\citet{Krimm2020GCN}; %200729A
\hypertarget{Dichiara2020GCN28270}{(59)}~\citet{Dichiara2020GCN28270}; %200729A
\hypertarget{Lu2022}{(60)}~\citet{Lu2022}; %211227A
\hypertarget{Cenko2022}{(61)}~\citet{Cenko2022}%220611A 
}
\end{deluxetable*}

%% file: Tab/sfr_luminosity.tex
\begin{ThreePartTable}
\begin{TableNotes}[para,flushleft]
\footnotesize
\item[] \textbf{References.}
\hypertarget{sfrtab:Bloom06}{(1)}~\citet{Bloom2006}; % 050509B
\hypertarget{sfrtab:Albareti17}{(2)}~\citet{Albareti2017}; % 050509B
\hypertarget{sfrtab:Bloom07}{(3)}~\citet{Bloom2007}; % 060502B
\hypertarget{sfrtab:NicuesaGuelbenzu21}{(4)}~\citet{NicuesaGuelbenzu2021}; % 080905A
\hypertarget{sfrtab:Rowlinson10}{(5)}~\citet{Rowlinson2010}; % 080905A
\hypertarget{sfrtab:Nugent22}{(6)}~\citet{Nugent2022}; % 130822A
\hypertarget{sfrtab:OConnor22}{(7)}~\citet{OConnor2022}; % 130822A
\hypertarget{sfrtab:Troja18}{(8)}~\citet{Troja2018}; % 150101B
\hypertarget{sfrtab:Dey19}{(9)}~\citet{Dey2019}; % 120422A, 130427A, 150818A, 161219B, 171205A, 060614, 060428B, 060912A, 061021, 080517, 090417B, 111005A, 130925A, 200716C, 200729A, 211227A, 220611A
\hypertarget{sfrtab:Troja19}{(10)}~\citet{Troja2019}; % 160821B
\hypertarget{sfrtab:Schroeder25}{(11)}~\citet{Schroeder2025}; % 231117A
\hypertarget{sfrtab:Flewelling20}{(12)}~\citet{Flewelling2020}; % 231117A, 060218, 191019A, 051109B
\hypertarget{sfrtab:Wiersema07}{(13)}~\citet{Wiersema2007}; % 060218
\hypertarget{sfrtab:Izzo17}{(14)}~\citet{Izzo2017}; % 100316D
\hypertarget{sfrtab:Schulze14}{(15)}~\citet{Schulze2014}; % 120422A
\hypertarget{sfrtab:Levesque12}{(16)}~\citet{Levesque2012}; % 120422A
\hypertarget{sfrtab:Xu13}{(17)}~\citet{Xu2013}; % 130427A
\hypertarget{sfrtab:Taggart21}{(18)}~\citet{Taggart2021}; % 150818A
\hypertarget{sfrtab:Cano17}{(19)}~\citet{Cano2017}; % 161219B
\hypertarget{sfrtab:Thone24}{(20)}~\citet{Thone2024}; % 171205A
\hypertarget{sfrtab:Rossi26}{(21)}~\citet{Rossi2026}; % 180728A
\hypertarget{sfrtab:Gupta22}{(22)}~\citet{Gupta2022}; % 190829A
\hypertarget{sfrtab:SanchezRamirez24}{(23)}~\citet{SanchezRamirez2024}; % 221009A
\hypertarget{sfrtab:Levan23_221009A}{(24)}~\citet{Levan2023_221009A}; % 221009A host photometry
\hypertarget{sfrtab:Berger05}{(25)}~\citet{Berger2005}; % 050724
\hypertarget{sfrtab:Gorosabel06}{(26)}~\citet{Gorosabel2006}; % 050724
\hypertarget{sfrtab:Thone14}{(27)}~\citet{Thone2014}; % 060505
\hypertarget{sfrtab:Fynbo06}{(28)}~\citet{Fynbo2006}; % 060614
\hypertarget{sfrtab:Levan23b}{(29)}~\citet{Levan2023}; % 191019A
\hypertarget{sfrtab:Troja22}{(30)}~\citet{Troja2022}; % 211211A
\hypertarget{sfrtab:Rossi14}{(31)}~\citet{Rossi2014}; % 050219A
\hypertarget{sfrtab:Mirabal07}{(32)}~\citet{Mirabal2007}; % 050826
\hypertarget{sfrtab:ThisWork}{(33)}~This work; % 051109B, 111225A, 150727A, 200716C, 200729A, 220611A
\hypertarget{sfrtab:Perley06}{(34)}~\citet{Perley2007AIP}; % 060428B
\hypertarget{sfrtab:Levan07}{(35)}~\citet{Levan2007}; % 060912A
\hypertarget{sfrtab:Kruhler15}{(36)}~\citet{Kruhler2015}; % 061021
\hypertarget{sfrtab:Stanway15}{(37)}~\citet{Stanway2015}; % 080517
\hypertarget{sfrtab:Niino17}{(38)}~\citet{Niino2017}; % 090417B, 111225A (FOCAS data)
\hypertarget{sfrtab:Michalowski18}{(39)}~\citet{Michalowski2018}; % 111005A
\hypertarget{sfrtab:Schady15}{(40)}~\citet{Schady2015}; % 130925A
 \hypertarget{sfrtab:Kennicutt08}{(41)}~\citet{Kennicutt2008}; % 200729A (NGC 4242; GALEX Local Volume survey)
\hypertarget{sfrtab:Ferro23}{(42)}~\citet{Ferro2023}. % 211227A
\end{TableNotes}
\setlength{\LTcapwidth}{\textwidth}
\setlength{\LTleft}{0pt}
\setlength{\LTright}{0pt}
\renewcommand{\TPTminimum}{\linewidth}
\begin{longtable}{@{\extracolsep{\fill}}lllrrl}
\caption{Host-galaxy SFRs and rest-frame $B$-band luminosity for GRBs at $z<0.35$. Apparent magnitudes are AB and corrected for Milky Way extinction. 
SFR values are normalized to a Chabrier IMF.} \label{tab:sfr}\\
\hline
GRB & SFR & Photometry & $M_{B}$ & $L_B/L_B^*$ & Ref.\\
 & ($M_\odot\,\mathrm{yr}^{-1}$) & (AB mag) & (mag) & & \\
\midrule
\endfirsthead
\multicolumn{6}{c}{\tablename\ \thetable{} -- \textit{Continued from previous page}} \\
\midrule
GRB & SFR & Photometry & $M_{B}$ & $L_B/L_B^*$ & Ref.\\
 & ($M_\odot\,\mathrm{yr}^{-1}$) & (AB mag) & (mag) & & \\
\midrule
\endhead
\midrule
\multicolumn{6}{r}{\textit{Next page}} \\
\endfoot
\bottomrule
\insertTableNotes
\endlastfoot
\multicolumn{6}{@{}l}{\textbf{SGRBs}}\\
050509B & $<0.06$ & $g=18.48\pm0.02$ & $-22.2$ & $4.8$ & \hyperlink{sfrtab:Bloom06}{1}, \hyperlink{sfrtab:Albareti17}{2} \\
 & & $r=17.07\pm0.01$ & & & \\  %OK
060502B & $\sim0.5$ & $g^\prime=20.12\pm0.03$ & $-21.2$ & $1.9$ & \hyperlink{sfrtab:Bloom07}{3} \\
  & & $R=18.83\pm0.01$ & & & \\ %OK.
080905A & $\sim1.5$ & -- & $-20.4$ & $0.9$ & \hyperlink{sfrtab:NicuesaGuelbenzu21}{4}, \hyperlink{sfrtab:Rowlinson10}{5} \\  %OK
130822A & $0.30^{+0.05}_{-0.04}$ & $G=18.76\pm0.03$ & $-20.6$ & $1.1$ & \hyperlink{sfrtab:Nugent22}{6}, \hyperlink{sfrtab:OConnor22}{7} \\
 & & $R=18.13\pm0.03$ & & & \\ %OK
150101B & $\lesssim 0.5$ & $g=17.095\pm0.001$ & $-21.9$ & $3.8$ & \hyperlink{sfrtab:Troja18}{8}, \hyperlink{sfrtab:Dey19}{9} \\
 & & $r=16.135\pm0.002$ & & & \\ %OK
160821B & $\sim0.9$ & --- & $-19.9$ & $0.6$ & \hyperlink{sfrtab:Troja19}{10} \\ %OK
231117A & $0.39^{+0.04}_{-0.03}$ & $g=21.94\pm0.08$ & $-18.6$ & $0.17$ & \hyperlink{sfrtab:Schroeder25}{11}, \hyperlink{sfrtab:Flewelling20}{12} \\
 & & $r=21.59\pm0.07$ & & & \\ %OK
\addlinespace[6pt]
\multicolumn{6}{@{}l}{\textbf{Group A}}\\
060218 & $0.041\pm 0.005$ & $g=20.13\pm0.03$ & $-15.7$ & $0.011$ & \hyperlink{sfrtab:Flewelling20}{12}, \hyperlink{sfrtab:Wiersema07}{13}\\
 & & $r=19.80\pm0.03$ & & & \\ %OK
100316D & $1.20\pm 0.08$ & --- & $-18.5$  & $0.16$ & \hyperlink{sfrtab:Izzo17}{14} \\
120422A & $0.8–1.7$ & $g=21.61\pm0.02$ & $-19.3$ & $0.33$ &
\hyperlink{sfrtab:Dey19}{9}, \hyperlink{sfrtab:Schulze14}{15}, \hyperlink{sfrtab:Levesque12}{16}\\
 & & $r=20.98\pm0.02$ & & & \\ %OK
130427A & $2^{+5}_{-1}$ & $g=21.91\pm0.03$ & $-19.6$ & $0.44$ &  \hyperlink{sfrtab:Dey19}{9}, \hyperlink{sfrtab:Xu13}{17} \\
 & & $r=21.25\pm0.03$ & & & \\ %OK
150818A & $1.5^{+0.9}_{-0.6}$ & $g=22.26\pm0.05$ & $-18.6$ & $0.18$ & \hyperlink{sfrtab:Dey19}{9}, \hyperlink{sfrtab:Taggart21}{18} \\
 & & $r=21.68\pm0.06$ & & & \\
161219B & $0.25^{+0.30}_{-0.17}$ & $g=20.98\pm0.02$ & $-18.2$ & $0.12$ & \hyperlink{sfrtab:Dey19}{9}, \hyperlink{sfrtab:Cano17}{19} \\
 & & $r=20.47\pm0.02$ & & & \\
171205A & $1.44\pm 0.04$ & $g=15.7410\pm0.0007$ & $-20.2$ & $0.74$ & \hyperlink{sfrtab:Dey19}{9}, \hyperlink{sfrtab:Thone24}{20} \\
 & & $r=15.1763\pm0.0006$ & & & \\ %OK
180728A & $0.28^{+0.31}_{-0.27}$ & --- & $-16.0$ & $0.016$ &  \hyperlink{sfrtab:Rossi26}{21} \\ %OK
190829A & $6.87$ & $g=16.2750\pm0.0008$ & $-21.4$ & $2.25$ & \hyperlink{sfrtab:Dey19}{9}, \hyperlink{sfrtab:Gupta22}{22} \\
 & & $r=15.3567\pm0.0006$ & & & \\ %OK
221009A & $0.8$ & $F625W=21.40\pm0.08$ & $-17.6$ & $0.07$ & \hyperlink{sfrtab:SanchezRamirez24}{23}, \hyperlink{sfrtab:Levan23_221009A}{24}\\
 & & $F775W=21.27\pm0.14$ & & & \\ % OK Levan+2023 (ApJL 946 L28) Sec. 5 GALFITM host model mags F625W 24.88+-0.08, F775W 23.80+-0.14, minus A = E(B-V)_SFD x SF11 WFC3 coeff (2.259, 1.643), E(B-V)_SFD = 1.539 -> A 3.48, 2.53. 
\addlinespace[6pt]
\multicolumn{6}{@{}l}{\textbf{Group B}}\\
050724 & $<0.02$ & $V=19.07\pm0.05$ & $-21.3$ & $2.1$ & \hyperlink{sfrtab:Berger05}{25}, \hyperlink{sfrtab:Gorosabel06}{26} \\
 & & $R=18.55\pm0.03$ & & & \\  % OK
060505 & $\sim0.8$ & $B=18.76\pm0.02$ & $-19.4$ & $0.4$ & \hyperlink{sfrtab:Thone14}{27} \\ 
 & & $V=18.27\pm0.02$ & & & \\  % OK
060614 & $\sim 0.009$ & $g=23.01\pm0.04$ & $-15.8$ & $0.013$ & \hyperlink{sfrtab:Dey19}{9}, \hyperlink{sfrtab:Fynbo06}{28} \\
 & & $r=22.46\pm0.03$ & & & \\  % OK
191019A & $0.06^{+0.08}_{-0.03}$ & $g=20.05\pm0.02$ & $-20.7$ & $1.20$ & \hyperlink{sfrtab:Flewelling20}{12}, \hyperlink{sfrtab:Levan23b}{29} \\
 & & $r=19.093\pm0.009$ & & & \\ % OK
211211A & $0.06\pm 0.02$ & $g=19.93\pm0.01$ & $-17.7$ & $0.08$ & \hyperlink{sfrtab:Troja22}{30}\\
 & & $r=19.50\pm0.01$  & & & \\ % OK
\addlinespace[6pt]
\multicolumn{6}{@{}l}{\textbf{Group C}}\\
050219A & $0.06^{+0.01}_{-0.02}$ & $g=20.85\pm0.04$ & $-19.6$ & $0.42$ & \hyperlink{sfrtab:Rossi14}{31} \\
 & & $r=19.76\pm0.04$ & & & \\ % OK
050826 & $2\pm0.9$ & --- & $-19.8$ & $0.54$ & \hyperlink{sfrtab:Mirabal07}{32} \\ % OK
051109B & $0.08$ & $g=16.905\pm0.005$ & $-20.8$ & $1.28$ & \hyperlink{sfrtab:Flewelling20}{12}, \hyperlink{sfrtab:ThisWork}{33} \\
 & & $r=16.075\pm0.003$ & & & \\ % OK
060428B & $0.19\pm 0.05$ & $g=21.30\pm0.03$ & $-21.0$ & $1.57$ & \hyperlink{sfrtab:Dey19}{9}, \hyperlink{sfrtab:Perley06}{34} \\
 & & $r=19.72\pm0.01$ & & & \\ % OK
060912A & $<0.02$ & $g=17.081\pm0.001$ & $-21.0$ & $1.61$ & \hyperlink{sfrtab:Dey19}{9}, \hyperlink{sfrtab:Levan07}{35} \\
 & & $r=16.1122\pm0.0009$ & & & \\ % OK
061021 & $0.05^{+0.03}_{-0.01}$ & $g=25.1\pm0.3$ & $-16.3$ & $0.02$ & \hyperlink{sfrtab:Dey19}{9}, \hyperlink{sfrtab:Kruhler15}{36} \\
 & & $r=24.6\pm0.2$ & & & \\ % OK
080517 & $\sim16$ & $g=17.980\pm0.001$ & $-20.0$ & $0.63$ & \hyperlink{sfrtab:Dey19}{9}, \hyperlink{sfrtab:Stanway15}{37} \\
 & & $r=17.579\pm0.002$ & & & \\ % OK
090417B & $1.3^{+0.5}_{-0.3}$ & $g=22.51\pm0.05$ & $-19.2$ & $0.31$ & \hyperlink{sfrtab:Dey19}{9}, \hyperlink{sfrtab:Niino17}{38} \\
 & & $r=21.59\pm0.04$ & & & \\ % OK
111005A & $0.42^{+0.06}_{-0.05}$ & $g=14.7270\pm0.0006$ & $-18.9$ & $0.23$ & \hyperlink{sfrtab:Dey19}{9}, \hyperlink{sfrtab:Michalowski18}{39} \\
 & & $r=14.1347\pm0.0004$ & & & \\ % OK
\modify{111225A} & \modify{$0.07-0.26$} & \modify{$V=23.87\pm0.21$} & \modify{$-16.9$} & \modify{$0.034$} & \hyperlink{sfrtab:ThisWork}{33}, \hyperlink{sfrtab:Niino17}{38} \\ % SFR: Keck/LRIS H-alpha (4.2 sigma), Galactic- and slit-loss-corrected, KE12, no internal-extinction correction (lower bound in practice); photometry: Subaru/FOCAS (Niino+2017 data) reduced in this work, Galactic-corrected
 & & \modify{$R=23.56\pm0.14$} & & & \\
130925A & $2.4^{+1.6}_{-1.3}$ & $g=22.80\pm0.03$ & $-18.9$ & $0.24$ & \hyperlink{sfrtab:Dey19}{9}, \hyperlink{sfrtab:Schady15}{40} \\
 & & $r=21.88\pm0.02$ & & & \\
\modify{150727A} & \modify{$0.34^{+0.26}_{-0.13}$} & \modify{$g^\prime=25.4\pm0.3$} &\modify{$-16.1$} &\modify{$0.016$} & \hyperlink{sfrtab:ThisWork}{33} \\ % SFR: XSGRB X-shooter H-alpha, Balmer-decrement (Cardelli) internal-extinction correction, KE12; photometry: GROND 2017 stacks (this work), Galactic-corrected
 & & \modify{$r^\prime=24.5\pm0.2$} & & & \\
200716C & $0.2$ & $g=20.619\pm0.006$ & $-21.1$ & $1.79$ & \hyperlink{sfrtab:Dey19}{9}, \hyperlink{sfrtab:ThisWork}{33} \\
 & & $r=19.617\pm0.004$ & & & \\
200729A & $\sim0.04$ & $g=11.4759\pm0.0002$ & $-17.8$ & $0.08$ & \hyperlink{sfrtab:Dey19}{9}, \hyperlink{sfrtab:ThisWork}{33}, \hyperlink{sfrtab:Kennicutt08}{41}\\
 & & $r=10.9773\pm0.0002$ & & & \\
211227A & $0.42\pm 0.02$ & $g=20.056\pm0.009$ & $-20.5$ & $1.00$ & \hyperlink{sfrtab:Dey19}{9}, \hyperlink{sfrtab:Ferro23}{42} \\
 & & $r=18.972\pm0.006$ & & & \\ % OK
220611A & $<0.05$ & $g=15.2027\pm0.0003$ & $-21.3$ & $2.14$ & \hyperlink{sfrtab:Dey19}{9}, \hyperlink{sfrtab:ThisWork}{33} \\
 & & $r=14.3430\pm0.0002$ & & & \\ % OK
\end{longtable}
\end{ThreePartTable}

%% file: Tab/constraints.tex
\begin{ThreePartTable}
\begin{TableNotes}[para,flushleft]
\footnotesize
\item[] \textbf{Notes.}\\
\item[a] Rest-frame epoch earlier than 1.9\,d, prior to the emergence of the SN component in the SN\,1998bw template. $\Delta M$ is computed against the 1.9\,d template magnitude and is therefore conservative.\\
\item[b] Rest-frame effective wavelength falls outside the template coverage and the template is extrapolated there.\\
\item[c] Host-contaminated measurement 
% (GRB\,130925A, \citealt{Greiner2014}; GRB\,200716C, \citealt{GCN28151}) 
is used as an upper limit on the transient flux.
%\item[] $\approx$ marks a measurement published without an uncertainty.
\item[] \\
\textbf{References.}
\hypertarget{ctab:Hjorth2005}{(1)}~\citet{Hjorth2005}; % 050509B
\hypertarget{ctab:Perley2011}{(2)}~\citet{Perley2011}; % 060502B, 070809, 051109B, 060428B
\hypertarget{ctab:Rowlinson2010}{(3)}~\citet{Rowlinson2010}; % 080905A
\hypertarget{ctab:Rastinejad2021}{(4)}~\citet{Rastinejad2021}; % 130822A
\hypertarget{ctab:Troja2018}{(5)}~\citet{Troja2018}; % 150101B
% \hypertarget{ctab:Fong2015}{(5)}~\citet{Fong2015}; % 150101B
\hypertarget{ctab:Troja2019}{(6)}~\citet{Troja2019}; % 160821B
\hypertarget{ctab:GCN30983}{(7)}~\citet{GCN30983}; % 210919A
\hypertarget{ctab:Anderson2025}{(8)}~\citet{Anderson2025}; % 231117A
\hypertarget{ctab:Malesani2007}{(9)}~\citet{Malesani2007}; % 050724A
\hypertarget{ctab:Ofek2007}{(10)}~\citet{Ofek2007}; % 060505
\hypertarget{ctab:Yang2015}{(11)}~\citet{Yang2015}; % 060614
\hypertarget{ctab:Stratta2025}{(12)}~\citet{Stratta2025}; % 191019A
\hypertarget{ctab:Troja2022}{(13)}~\citet{Troja2022}; % 211211A
\hypertarget{ctab:Mirabal2007}{(14)}~\citet{Mirabal2007}; % 050826
\hypertarget{ctab:Roming2017}{(15)}~\citet{Roming2017}; % 060912A, 061021, 080517
\hypertarget{ctab:Michalowski2018}{(16)}~\citet{Michalowski2018}; % 111005A
\hypertarget{ctab:GCN12793}{(17)}~\citet{GCN12793}; % 111225A
\hypertarget{ctab:Greiner2014}{(18)}~\citet{Greiner2014}; % 130925A
\hypertarget{ctab:GCN18095}{(19)}~\citet{GCN18095}; % 150727A
\hypertarget{ctab:GCN28151}{(20)}~\citet{GCN28151}; % 200716C
\hypertarget{ctab:Dichiara2020GCN28270}{(21)}~\citet{Dichiara2020GCN28270}; % 200729A
\hypertarget{ctab:thiswork}{(22)}~This work; % 211227A
\hypertarget{ctab:Malesani2022gcn32222}{(23)}~\citet{Malesani2022gcn32222}; % 220611A
\hypertarget{ctab:Cenko2005GCN3409}{(24)}~\citet{Cenko2005GCN3409}; % 050509B
\hypertarget{ctab:GCN5077}{(25)}~\citet{GCN5077}; % 060502B
\hypertarget{ctab:Jin2020}{(26)}~\citet{Jin2020}; % 070809
\hypertarget{ctab:NicuesaGuelbenzu2012}{(27)}~\citet{NicuesaGuelbenzu2012}; % 100628A
\hypertarget{ctab:GCN15121}{(28)}~\citet{GCN15121}; % 130822A
\hypertarget{ctab:Fong2016}{(29)}~\citet{Fong2016}; % 150101B
\hypertarget{ctab:GCN31567}{(30)}~\citet{GCN31567}; % 210919A
\hypertarget{ctab:Levan2021GCN31235}{(31)}~\citet{Levan2021GCN31235}; % 211211A
\hypertarget{ctab:GCN3048}{(32)}~\citet{GCN3048}; % 050219A
\hypertarget{ctab:Levan2026}{(33)}~\citet{Levan2026}; % 051109B
\hypertarget{ctab:GCN5034}{(34)}~\citet{GCN5034}; % 060428B
\hypertarget{ctab:GCN5747}{(35)}~\citet{GCN5747}; % 061021
\hypertarget{ctab:Tanga2018}{(36)}~\citet{Tanga2018}; % 111005A
\hypertarget{ctab:Cenko2011GCN12733}{(37)}~\citet{Cenko2011GCN12733}; % 111225A
\hypertarget{ctab:GCN28180}{(38)}~\citet{GCN28180}; % 200729A
\hypertarget{ctab:Ferro2023}{(39)}~\citet{Ferro2023}; % 211227A
\hypertarget{ctab:OConnor2022gcn32228}{(40)}~\citet{OConnor2022gcn32228}. % 220611A
\end{TableNotes}
\setlength{\LTcapwidth}{\textwidth}
\setlength{\LTleft}{0pt}
\setlength{\LTright}{0pt}
\renewcommand{\TPTminimum}{\linewidth}
\begin{longtable}{@{\extracolsep{\fill}}lccccccrc}
\caption{Supernova and kilonova constraints for each burst. $\Delta M$ measures the difference between the observation and the template light curve at the same rest-frame epoch $t_{\rm rest}$ and rest-frame wavelength $\lambda_{\rm rest}$. Lower limits on $\Delta M$ indicate a non-detection or a measurement dominated by host-galaxy light (note $^{\rm c}$). $t_{\rm obs}$ is the observed epoch of that measurement, and $m_{\rm obs}$ is in AB magnitudes and corrected for Galactic extinction with the \citet{Schlafly2011} coefficients. Values published in the Vega system have been converted into AB using the offsets of \citet{Blanton2007} and, for Swift/UVOT, of \citet{Breeveld2011}.} \label{tab:constraints}\\
\hline
GRB & $z$ & Telescope/Filter & $t_{\rm obs}$ & $t_{\rm rest}$ & $\lambda_{\rm rest}$ & $m_{\rm obs}$ & $\Delta M$ & Ref.\\
 & & & (d) & (d) & ($\mu$m) & (AB) & (mag) & \\
\midrule
\endfirsthead
\multicolumn{9}{c}{\tablename\ \thetable{} -- \textit{Continued from previous page}} \\
\midrule
GRB & $z$ & Telescope/Filter & $t_{\rm obs}$ & $t_{\rm rest}$ & $\lambda_{\rm rest}$ & $m_{\rm obs}$ & $\Delta M$ & Ref.\\
 & & & (d) & (d) & ($\mu$m) & (AB) & (mag) & \\
\midrule
\endhead
\midrule
\multicolumn{9}{r}{\textit{Next page}} \\
\endfoot
\bottomrule
\insertTableNotes
\endlastfoot
\multicolumn{9}{l}{\textit{Supernova Constraints: comparison with SN\,1998bw}}\\
\hline
\multicolumn{9}{@{}l}{\textbf{SGRBs}}\\
050509B & 0.225 & VLT/\textit{V} & 23 & 19 & 0.45 & $>26.4$ & $>5.2$ & \hyperlink{ctab:Hjorth2005}{1}\\
060502B & 0.287 & Keck/\textit{R} & 28 & 22 & 0.49 & $>25.7$ & $>4.0$ & \hyperlink{ctab:Perley2011}{2}\\
070809 & 0.2187 & Keck/\textit{g} & 187 & 153 & 0.39 & $>26.8$ & $>1.3$ & \hyperlink{ctab:Perley2011}{2}\\
080905A & 0.1218 & VLT/\textit{R} & 1.5 & 1.4\tnote{a} & 0.57 & $>25.2$ & $>3.6$ & \hyperlink{ctab:Rowlinson2010}{3}\\
130822A & 0.154 & UKIRT/\textit{H} & 5.7 & 4.9 & 1.4\tnote{b} & $>22.8$ & $>0.9$ & \hyperlink{ctab:Rastinejad2021}{4}\\
150101B & 0.134 & HST/\textit{F606W} & 41 & 36 & 0.52 & $>25.6$ & $>4.6$ & \hyperlink{ctab:Troja2018}{5}\\
160821B & 0.161 & HST/\textit{F606W} & 10 & 8.9 & 0.51 & $27.9\pm0.3$ & $7.4$ & \hyperlink{ctab:Troja2019}{6}\\
210919A & 0.2415 & VLT/\textit{Ic} & 13 & 11 & 0.64 & $>25.6$ & $>4.3$ & \hyperlink{ctab:GCN30983}{7}\\
231117A & 0.257 & NTT/\textit{R} & 6.9 & 5.5 & 0.51 & $23.4\pm0.3$ & $1.4$ & \hyperlink{ctab:Anderson2025}{8}\\
\addlinespace[6pt]
\multicolumn{9}{@{}l}{\textbf{Group B}}\\
050724A & 0.257 & VLT/\textit{R} & 32 & 26 & 0.51 & $>24.6$ & $>2.9$ & \hyperlink{ctab:Malesani2007}{9}\\
060505 & 0.089 & HST/\textit{F814W} & 14 & 13 & 0.74 & $>27.3$ & $>8.0$ & \hyperlink{ctab:Ofek2007}{10}\\
060614 & 0.125 & HST/\textit{F814W} & 31 & 28 & 0.71 & $>27.7$ & $>7.6$ & \hyperlink{ctab:Yang2015}{11}\\
191019A & 0.248 & Gemini/\textit{r} & 11 & 9.1 & 0.50 & $>25.7$ & $>4.3$ & \hyperlink{ctab:Stratta2025}{12}\\
211211A & 0.0763 & CAHA/\textit{r} & 21 & 19 & 0.58 & $>24.4$ & $>6.0$ & \hyperlink{ctab:Troja2022}{13}\\
\addlinespace[6pt]
\multicolumn{9}{@{}l}{\textbf{Group C}}\\
050826 & 0.296 & MDM/\textit{R} & 2.2 & 1.7\tnote{a} & 0.49 & $>21.6$ & $>-1.7$ & \hyperlink{ctab:Mirabal2007}{14}\\
051109B & 0.08 & Keck/\textit{R} & 258 & 239 & 0.59 & $>26.4$ & $>2.8$ & \hyperlink{ctab:Perley2011}{2}\\
060428B & 0.348 & Keck/\textit{R} & 32 & 24 & 0.47 & $>25.2$ & $>2.8$ & \hyperlink{ctab:Perley2011}{2}\\
060912A & 0.0939 & UVOT/\textit{u} & 9.8 & 9.0 & 0.32\tnote{b} & $>22.5$ & $>1.1$ & \hyperlink{ctab:Roming2017}{15}\\
061021 & 0.3463 & UVOT/\textit{u} & 23 & 17 & 0.26\tnote{b} & $>23.6$ & $>-1.6$ & \hyperlink{ctab:Roming2017}{15}\\
% % 090417B & 0.345 & UVOT/\textit{u} & 13 & 10.0 & 0.26\tnote{b} & $>23.2$ & $>-30.6$ & \hyperlink{ctab:Roming2017}{15}\\   % no constraint: host A_V is a lower limit
111005A & 0.01326 & Spitzer/\textit{IRAC1} & 9.2 & 9.0 & 3.5\tnote{b} & $>21.9$ & $>5.1$ & \hyperlink{ctab:Michalowski2018}{16}\\
111225A & 0.297 & CrAO/\textit{R} & 3.8 & 2.9 & 0.49 & $22.1\pm0.1$ & $-0.9$ & \hyperlink{ctab:GCN12793}{17}\\
130925A & 0.347 & VLT/\textit{H} & 18 & 13 & 1.2\tnote{b} & $>20.5$\tnote{c} & $>-3.3$ & \hyperlink{ctab:Greiner2014}{18}\\
150727A & 0.313 & OAN/\textit{i} & 1.4 & 1.0\tnote{a} & 0.58 & $>22.5$ & $>-1.1$ & \hyperlink{ctab:GCN18095}{19}\\
200716C & 0.341 & Assy/\textit{r} & 2.8 & 2.1 & 0.46 & $>19.3$\tnote{c} & $>-4.2$ & \hyperlink{ctab:GCN28151}{20}\\
200729A & 0.00176 & LDT/\textit{i} & 18 & 18 & 0.76 & $>21.8$ & $>11.0$ & \hyperlink{ctab:Dichiara2020GCN28270}{21}\\
211227A & 0.228 & Gemini/\textit{i} & 12 & 10 & 0.62 & $>25.7$ & $>4.5$ & \hyperlink{ctab:thiswork}{22}\\
220611A & 0.049 & VLT/\textit{I} & 6.7 & 6.4 & 0.84\tnote{b} & $23.0\pm0.2$ & $4.3$ & \hyperlink{ctab:Malesani2022gcn32222}{23}\\
\hline
\multicolumn{9}{l}{\textit{Kilonova Constraints: comparison with AT2017gfo}}\\
\hline
\multicolumn{9}{@{}l}{\textbf{SGRBs}}\\
050509B & 0.225 & Keck/\textit{R} & 1.1 & 0.9 & 0.52 & $>25.7$ & $>1.5$ & \hyperlink{ctab:Cenko2005GCN3409}{24}\\
060502B & 0.287 & Gemini/\textit{R} & 0.7 & 0.5 & 0.49 & $>24.4$ & $>-0.2$ & \hyperlink{ctab:GCN5077}{25}\\
070809 & 0.2187 & Keck/\textit{R} & 1.5 & 1.2 & 0.52 & $>25.1$ & $>0.5$ & \hyperlink{ctab:Jin2020}{26}\\
080905A & 0.1218 & VLT/\textit{R} & 0.6 & 0.5 & 0.57 & $24.4\pm0.3$ & $1.8$ & \hyperlink{ctab:Rowlinson2010}{3}\\
100628A & 0.102 & GROND/\textit{r} & 0.7 & 0.7 & 0.56 & $>24.5$ & $>2.3$ & \hyperlink{ctab:NicuesaGuelbenzu2012}{27}\\
130822A & 0.154 & Gemini/\textit{i} & 0.9 & 0.8 & 0.66 & $>24.5$ & $>1.1$ & \hyperlink{ctab:GCN15121}{28}\\
150101B & 0.134 & Magellan/\textit{r} & 1.7 & 1.5 & 0.55 & $23.0\pm0.2$ & $-0.8$ & \hyperlink{ctab:Fong2016}{29}\\
160821B & 0.161 & HST/\textit{F160W} & 3.7 & 3.2 & 1.3 & $24.5\pm0.1$ & $0.5$ & \hyperlink{ctab:Troja2019}{6}\\
210919A & 0.2415 & Maidanak/\textit{R} & 0.9 & 0.8 & 0.51 & $>24.1$ & $>-0.2$ & \hyperlink{ctab:GCN31567}{30}\\
231117A & 0.257 & VLT/\textit{Ks} & 6.9 & 5.5 & 1.7 & $>23.8$ & $>-1.7$ & \hyperlink{ctab:Anderson2025}{8}\\
\addlinespace[6pt]
\multicolumn{9}{@{}l}{\textbf{Group B}}\\
050724A & 0.257 & VLT/\textit{I} & 3.5 & 2.8 & 0.64 & $25.1\pm0.3$ & $-1.1$ & \hyperlink{ctab:Malesani2007}{9}\\
060505 & 0.089 & Gemini/\textit{g} & 1.1 & 1.0 & 0.44 & $22.4\pm0.1$ & $-0.3$ & \hyperlink{ctab:Ofek2007}{10}\\
060614 & 0.125 & VLT/\textit{I} & 2.8 & 2.5 & 0.71 & $22.4\pm0.2$ & $-1.8$ & \hyperlink{ctab:Yang2015}{11}\\
191019A & 0.248 & GROND/\textit{i} & 1.5 & 1.2 & 0.61 & $23.4\pm0.3$ & $-1.4$ & \hyperlink{ctab:Stratta2025}{12}\\
211211A & 0.0763 & Gemini/\textit{K} & 4.0 & 3.7 & 2.0 & $22.4\pm0.1$ & $0.2$ & \hyperlink{ctab:Levan2021GCN31235}{31}\\
\addlinespace[6pt]
\multicolumn{9}{@{}l}{\textbf{Group C}}\\
050219A & 0.211 & LCO/\textit{I} & 0.7 & 0.6 & 0.66 & $>21.7$ & $>-2.3$ & \hyperlink{ctab:GCN3048}{32}\\
050826 & 0.296 & MDM/\textit{R} & 1.2 & 0.9 & 0.49 & $>21.6$ & $>-3.3$ & \hyperlink{ctab:Mirabal2007}{14}\\
051109B & 0.08 & CRTS/\textit{V} & 0.7 & 0.7 & 0.51 & $>20.1$ & $>-1.7$ & \hyperlink{ctab:Levan2026}{33}\\
060428B & 0.348 & MDM/\textit{R} & 1.0 & 0.7 & 0.47 & $22.6\pm0.3$ & $-2.5$ & \hyperlink{ctab:GCN5034}{34}\\
060912A & 0.0939 & UVOT/\textit{u} & 0.7 & 0.6 & 0.32 & $>21.0$ & $>-2.4$ & \hyperlink{ctab:Roming2017}{15}\\
061021 & 0.3463 & VLT/\textit{R} & 0.7 & 0.5 & 0.47 & $20.6\pm0.3$ & $-4.4$ & \hyperlink{ctab:GCN5747}{35}\\
080517 & 0.089 & UVOT/\textit{u} & 0.8 & 0.7 & 0.32 & $>19.0$ & $>-4.1$ & \hyperlink{ctab:Roming2017}{15}\\
% % 090417B & 0.345 & UVOT/\textit{u} & 1.5 & 1.1 & 0.26 & $>23.0$ & $>-33.8$ & \hyperlink{ctab:Roming2017}{15}\\   % no constraint: host A_V is a lower limit
111005A & 0.01326 & VLT/\textit{K} & 1.6 & 1.6 & 2.1 & $>21.8$ & $>3.0$ & \hyperlink{ctab:Tanga2018}{36}\\
111225A & 0.297 & Keck/\textit{R} & 1.0 & 0.8 & 0.49 & $22.8\pm0.3$ & $-2.0$ & \hyperlink{ctab:Cenko2011GCN12733}{37}\\
130925A & 0.347 & GROND/\textit{Ks} & 4.1 & 3.1 & 1.6 & $>20.3$\tnote{c} & $>-6.2$ & \hyperlink{ctab:Greiner2014}{18}\\
150727A & 0.313 & OAN/\textit{i} & 1.4 & 1.0 & 0.58 & $>22.5$ & $>-2.7$ & \hyperlink{ctab:GCN18095}{19}\\
200716C & 0.341 & CrAO/\textit{R} & 0.9 & 0.6 & 0.47 & $>19.4$\tnote{c} & $>-5.6$ & \hyperlink{ctab:GCN28151}{20}\\
200729A & 0.00176 & Nanshan/\textit{r} & 0.8 & 0.8 & 0.62 & $>20.2$ & $>6.6$ & \hyperlink{ctab:GCN28180}{38}\\
211227A & 0.228 & VLT/\textit{I} & 1.2 & 1.0 & 0.72 & $>25.8$ & $>1.4$ & \hyperlink{ctab:Ferro2023}{39}\\
220611A & 0.049 & Gemini/\textit{Ks} & 7.7 & 7.3 & 2.1 & $\approx21.8$ & $-0.1$ & \hyperlink{ctab:OConnor2022gcn32228}{40}\\
\end{longtable}
\end{ThreePartTable}

%% file: Tab/selection.tex
\begin{deluxetable}{lcl}
\ifdefined\rotateonfalse\rotateonfalse\fi
\tablecaption{\modify{Selection of SN-less LGRBs adopted to derive the rate of non-collapsar LGRBs (Sect.~\ref{sec:rate}).
We included LGRBs with ${\rm SFR}<1$ \,$M_\odot\,{\rm yr}^{-1}$ (low SFR)
and ${\rm sSFR}={\rm SFR}/L_B<3$\,$M_\odot\,{L_B^*}^{-1}\,{\rm yr}^{-1}$ (low sSFR).
Excluded events are marked Excl. These are bursts above these thresholds (high SFR and/or high sSFR), excluded from the sample of candidate non-collapsars, and those displaying a galactocentric offset larger than $10\arcsec$ (large offset), excluded from the rate calculation to minimize the impact of spurious host galaxy associations. Additionally, 
GRBs not observed through the standard BAT+XRT sequence (non-standard follow-up) were also excluded from the rate calculation.}
\label{tab:selection}}
\tabletypesize{\scriptsize}
\tablewidth{0pt}
\setlength{\tabcolsep}{15pt} 
\tablehead{
\colhead{GRB} &
\colhead{Sample} &
\colhead{Comment}
}
\startdata
\multicolumn{3}{l}{\textbf{Group B}}\\
\hline
050724A & Gold    & Low SFR, low sSFR \\
060505  & Excl.   & Non-standard follow-up \\
060614  & Gold    & Low SFR, low sSFR \\
191019A & Gold    & Low SFR, low sSFR \\
211211A & Gold    & Low SFR, low sSFR \\
\tableline
\multicolumn{3}{l}{\textbf{Group C}}\\
\hline
050219A & Silver  & Low SFR, low sSFR \\
050826  & Excl. & High SFR, high sSFR \\
050911  & Excl.   & Chance alignment \\
051109B & Excl.   & Large offset \\
060428B & Silver  & Low SFR, low sSFR \\
060912A & Excl.   & Large offset \\
061021  & Silver  & Low SFR, low sSFR \\
070412  & Excl.   & Chance alignment \\
070521  & Excl.   & Chance alignment \\
080517  & Excl.   & High SFR, high sSFR \\
090417B & Excl.   & High SFR, high sSFR \\
111005A & Excl.   & Non-standard follow-up \\
111225A & Excl.   & High SFR, high sSFR \\ 
130925A & Excl.   & High SFR, high sSFR \\
150424A & Excl.   & Chance alignment \\
150727A & Excl.   & Low SFR, high sSFR\\
200716C & Silver  & Low SFR, low sSFR \\
200729A & Excl.   & Large offset \\
211227A & Silver  & Low SFR, low sSFR \\
220611A & Excl.   & Large offset \\
\enddata
\end{deluxetable}